\documentclass[twocolumn,10pt]{IEEEtran}
\normalsize
\usepackage{bbm}
\usepackage{adjustbox}
\usepackage{cite,algorithm,algorithmic,amsmath,amssymb,amsthm,empheq,mhsetup}
\usepackage{subfigure,amsfonts,balance}
\usepackage{epstopdf}
\usepackage{setspace}
\usepackage[dvipsnames]{xcolor}
\DeclareMathOperator{\EEE}{\mathbb{E}}

\DeclareMathOperator{\C}{\mathbb{C}}

\DeclareMathOperator{\f}{\pmb{f}}

\DeclareMathOperator{\FF}{\mathcal{F}}

\DeclareMathOperator{\K}{\mathcal{K}}

\DeclareMathOperator{\OO}{\mathcal{O}}

\DeclareMathOperator{\vv}{\pmb{v}}
\DeclareMathOperator{\w}{\pmb{w}}

\DeclareMathOperator{\RRR}{\mathbb{R}}
\DeclareMathOperator{\RR}{\pmb{R}}

\DeclareMathOperator{\D}{\pmb{D}}

\DeclareMathOperator{\CN}{\mathcal{CN}}

\DeclareMathOperator{\MM}{\mathcal{M}}

\DeclareMathOperator{\NN}{\mathcal{N}}

\DeclareMathOperator{\DD}{\mathcal{D}}

\DeclareMathOperator{\x}{\pmb{x}}

\DeclareMathOperator{\y}{\pmb{y}}

\DeclareMathOperator{\uu}{\pmb{u}}

\DeclareMathOperator{\g}{\pmb{g}}

\DeclareMathOperator{\X}{\pmb{X}}

\DeclareMathOperator{\ETA}{\pmb{\eta}}

\DeclareMathOperator{\VARRHO}{\pmb{\varrho}}

\DeclareMathOperator{\pphi}{\pmb{\phi}}

\DeclareMathOperator{\VARPHI}{\pmb{\varphi}}

\DeclareMathOperator{\ALPHA}{\pmb{\alpha}}
\DeclareMathOperator{\BETA}{\pmb{\beta}}
\DeclareMathOperator{\CHI}{\pmb{\chi}}
\DeclareMathOperator{\ZETA}{\pmb{\zeta}}

\newtheorem{definition}{Definition}

\theoremstyle{remark}
\newtheorem{remark}{Remark}

\ifCLASSINFOpdf
\else
\fi

\begin{document}
\title{\Huge Cell-Free Massive MIMO\\ for Wireless Federated Learning}
\author{Tung~T.~Vu, Duy~T.~Ngo, Nguyen~H.~Tran, Hien~Quoc~Ngo, Minh~N.~Dao,
%, Duy~H.~N.~Nguyen
%~\IEEEmembership{Member,~IEEE,}
        and Richard~H.~Middleton
        %~\IEEEmembership{Fellow,~OSA,}
        %and~Jane~Doe,~\IEEEmembership{Life~Fellow,~IEEE}% <-this % stops a space
%\thanks{Manuscript received September 15, 2019.}
\thanks{Manuscript received December 18, 2019; revised April 21, 2020; accepted June 10, 2020. The editor coordinating the review of this paper and approving it for publication was Ming Xiao.}
\thanks{Tung T. Vu was supported by an ECR-HDR scholarship from The University of Newcastle. Duy T. Ngo was supported in part by the Australian Research Council Discovery Project (ARCDP) grant DP170100939, and in part by the Vietnam National Foundation for Science and Technology Development (NAFOSTED) under grant number 102.02-2018.320. Nguyen H. Tran was supported in part by the ARCDP grant DP200103718, and in part by NAFOSTED under grant number 102.02-2019.321. H. Q. Ngo was supported by the UK Research and Innovation Future Leaders Fellowships under Grant MR/S017666/1. Minh N. Dao was partially supported by ARCDP grants DP160101537 and DP190100555.}
\thanks{T.~T.~Vu, D.~T.~Ngo and R.~H.~Middleton are with the School of Electrical Engineering and Computing, The University of Newcastle, Callaghan, NSW 2308, Australia (e-mail: thanhtung.vu@uon.edu.au; \{duy.ngo, richard.middleton\}@newcastle.edu.au).}% <-this % stops a space
\thanks{N.~H.~Tran is with the School of Computer Science, The University of Sydney, Sydney, NSW 2006, Australia  (e-mail: nguyen.tran@sydney.edu.au).}
\thanks{H.~Q.~Ngo is with the School of Electronics, Electrical Engineering and Computer Science, Queen's University Belfast, Belfast BT7 1NN, United Kingdom (e-mail: hien.ngo@qub.ac.uk).}
\thanks{M.~N.~Dao is with the Department of Applied Mathematics, The University of New South Wales, Sydney, NSW 2052, Australia (e-mail: \mbox{daonminh@gmail.com}).}
%\thanks{D~H.~N.~Nguyen is with Department of Electrical and Computer
%Engineering, San Diego State University, San Diego,
%CA, USA 92182 (email: duy.nguyen@sdsu.edu).}
% <-this % stops a space
%\thanks{TCOM version based on Michael Shell's bare{\textunderscore}jrnl.tex version 1.3.}
}

\markboth{IEEE Transactions on Wireless Communications, VOL. XX, NO. X, XXX 2020}{IEEE Transactions on Wireless Communications, Accepted for Publication, 2020}



\maketitle
\vspace{-10mm}
\allowdisplaybreaks
\begin{abstract}
%\boldmath
This paper proposes a novel scheme for cell-free massive multiple-input multiple-output (CFmMIMO) networks to support any federated learning (FL) framework.
This scheme allows each instead of all the iterations of the FL framework to happen in a large-scale coherence time to guarantee a stable operation of an FL process.
To show how to optimize the FL performance using this proposed scheme, we consider an existing FL framework as an example and target FL training time minimization for this framework. An optimization problem is then formulated to jointly optimize the local accuracy, transmit power, data rate, and users' processing frequency.
This mixed-timescale stochastic nonconvex problem captures the complex interactions among the training time, and transmission and computation of training updates of one FL process.
By employing the online successive convex approximation approach, we develop a new algorithm to solve the formulated problem with proven convergence to the neighbourhood of its stationary points.
Our numerical results confirm that the presented joint design  reduces the training time by up to $55\%$ over baseline approaches.
They also show that CFmMIMO here requires the lowest training time for FL processes compared with cell-free time-division multiple access massive MIMO and collocated massive MIMO.
\end{abstract}
\vspace{-1mm}
\vspace{-2mm}

% \fi
\IEEEpeerreviewmaketitle

%\balance
\section{Introduction}
\label{sec:Introd}
The use of machine learning (ML) techniques in telecommunications industry has been growing dramatically in recent years \cite{letaief19CM,cala18CM}. One reason for this trend is the fast growing number of mobile devices, wearable devices and autonomous vehicles. They are generating a vast amount of data by using in-built sensors, e.g., microphones, GPS and camera, for critical applications such as traffic navigation, indoor localization, image recognition, natural language processing, and augmented reality
%\cite{dong19TMC,zhou18IOTJ}.
\cite{dong19TMC}.
In addition, the computational capabilities of these devices also grow significantly with dedicated hardware architecture and computing engines, e.g., the energy-efficient Qualcomm Hexagon Vector eXtensions on Snapdragon $835$ \cite{qualcomm17}. On-device artificial intelligence (AI) capabilities are predicted to be available on $80\%$ of all smartphones by $2022$ \cite{Gartner18}. It is therefore critical for telecommunications operators to start investigating into a future communications system that efficiently utilizes the empowered computation resources from mobile devices to solve  ML problems.

The typical ML framework used in the current telecommunications systems requires a cloud center to store and process raw data collected by the user equipment (UEs). However, such a centralized structure fails to support real-time applications because of its high latency \cite{zhu18arXiv}. The concept of mobile edge computing is introduced to process data at the edge nodes instead of the cloud center \cite{alam19JSAC,chen18JSAC}. Since the computational capability of mobile devices is growing noticeably, it is possible to even push the network computation further to the mobile device level \cite{ti19arXiv,tran19INFOCOM}. On the other hand, serious concerns about data privacy have recently been raised due to data being processed by third-party companies, e.g., Facebook, Apple. This urgently calls for a new class of ML frameworks that not only exploit the computational resources of the UEs for ML applications but also ensure data privacy.

\begin{figure}[t!]
\centering
\includegraphics[width=0.40\textwidth]{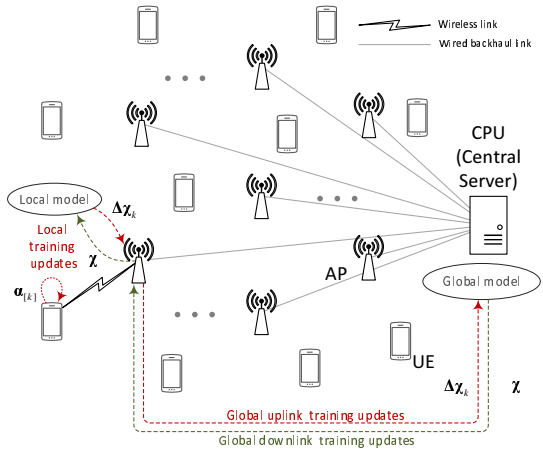}
\caption{Illustration of an FL process over communication networks and CFmMIMO network model used to support FL.}
\label{fig:systemmodel}
\end{figure}

A promising candidate for such ML frameworks is the recently developed Federated Learning (FL) \cite{ti19arXiv,mcmahan17}. As shown in Fig.~\ref{fig:systemmodel}, an FL process is an iterative process in which the UEs use their local training data to compute local model training updates, followed by sending the updates to a central server. The central server then aggregates these updates to compute the global training update, which is then sent back to the UEs to further assist their local update computation. This iterative process terminates when a certain learning accuracy level is attained.
Data privacy is protected by not sharing the local training \emph{data}, but only the local \emph{training updates} computed at the UEs using local computational resources.
%sharing only the local updates computed using local computational resource instead of sharing the local training data.
%Instead of sending the local training data, the UEs exploits their computational resource to compute the local update
%By not sharing the local training data and by computing the local updates directly at the UEs, the computational resource of UEs is exploited to protect the UEs' data privacy.
Uploading only the local training updates to the central server incurs a significantly lower delay, compared to uploading a large amount of raw data. This distributed approach facilitates a large-scale model training and more flexible data collection, albeit at the expense of UEs' computational resources \cite{tran19INFOCOM}.

With all the promising advantages listed above, FL has attracted much attention from both developer and researcher communities %\cite{hard18arXiv,bonawitz19arXiv,sattle19arXiv,sahu18arXiv,
%kang19arXiv,wang19JSAC,tran19INFOCOM,yang18arXiv,zeng19arXiv}.
\cite{hard18arXiv,sattle19arXiv,kang19arXiv,wang19JSAC,tran19INFOCOM,yang20TWC,chen19arXiv}.
In \cite{hard18arXiv}, an FL algorithm for keyboard prediction on smartphones is developed by Google.
%\cite{bonawitz19arXiv} builds a scalable production system for FL in real mobile devices using TensorFlow.
\cite{sattle19arXiv}
%and \cite{sahu18arXiv}
target improving the performance of the general FL algorithms, while \cite{kang19arXiv,wang19JSAC,yang20TWC,tran19INFOCOM,chen19arXiv} concentrate on optimizing the performance of FL algorithms used in wireless networks.
In particular,
\cite{sattle19arXiv} proposes a new compression framework for a communication-efficient FL system.
%The statistical heterogeneity issues of UEs are addressed in \cite{sahu18arXiv}.
\cite{tran19INFOCOM} aims to obtain the best trade-off between FL training time and UE energy consumption.
\cite{kang19arXiv} proposes an incentive mechanism that encourages the UEs with high-quality data to participate in FL systems. \cite{wang19JSAC} introduces a control algorithm to achieve the best trade-off between the number of local updates and that of global updates for a given resource budget. A joint device selection and beamforming design is proposed in \cite{yang20TWC} to enhance the performance of FL. %\cite{zeng19arXiv} proposes a bandwidth allocation and scheduling scheme that reduces the network energy consumption while guaranteeing performance for FL.
\cite{chen19arXiv} proposes a joint design of user selection, power control and subchannel allocation for minimizing the loss function of FL.

This paper investigates how to implement FL in a  wireless network. It is worth noting that the existing works \cite{tran19INFOCOM,yang20TWC} rely on an impractical assumption that the channel state information (CSI) remains unchanged during the whole FL process. In practice however, the channel changes in the order of milliseconds; and as such, certain system parameters for FL performance optimization, e.g., data rates and power control, would have become obsolete even before the FL process terminates.
In addition, it might not be most efficient to use orthogonal multiplexing approaches, e.g., orthogonal frequency-division multiple access (OFDMA) \cite{chen19arXiv} and time-division multiple access (TDMA) \cite{tran19INFOCOM}, for UEs to transmit their local updates. With a large number of UEs, the total training time could be significantly prolonged.
To both deal with the wireless channel dynamics and to serve UEs at the same time and in the same frequency bands, a new wireless network structure that supports FL is called for.

In this work, we propose using cell-free massive multiple-input multiple-output (CFmMIMO) \cite{zhang19A,ngo17TWC} for FL in a wireless environment. Here, a central processing unit (CPU) (i.e., the central server) is connected to a large number of access point (APs) via backhaul links. These APs then simultaneously serve UEs via wireless links using the same frequency bands with the CSI acquired via uplink (UL) training pilots.
%Recently, cell-free massive MIMO networks have attracted an increasing attention from the research community due to its noticeable benefits of providing handover-free and uniformly good services to all UEs \cite{zhang19A,bashar19TGCN,bashar19TWC,ngo17TWC}.
%In a cell-free massive MIMO network, UEs are simultaneously served by a large number of access points (APs) in the same frequency bands.
An important characteristic of massive MIMO is channel hardening \cite{ngo16}, i.e., the effective channel gain at the UEs is close to its expected value---a known deterministic constant \cite{ngo17TWC}.
%The achievable rates depends average effective channel gain instead of actual effective gain \cite{ngo17TWC}.
As such, the channels are reasonably stable during one large-scale coherence time $\widetilde{T}_{c}$\footnote{During one large-scale coherence time $\widetilde{T}_{c}$, the large-scale fading coefficients are reasonably invariant. The value of $\widetilde{T}_{c}$ can be empirically measured, in the same way for small-scale fading measurement. For indoor communications, the large-scale coherence time can be at least $40$ small-scale fading coherence time \cite{ngo17TWC}, and it has a time order of seconds.}. The channel dynamics due to small-scale fading thus have negligible effects on the FL processes.
%The achievable rate functions are also much simplified for further resource allocation \cite{vu19ICC}.
In addition, a CFmMIMO network also provides a high probability of coverage, making the FL processes less prone to the unfavorable UE links.

%Although CFmMIMO matches all the requirements to support FL, it still remains unclear of: (i) How to realize FL on a CFmMIMO network? (ii) Is the CFmMIMO network the best choice to support FL? This paper seeks the answers to these questions and makes the following research contributions.
Our research contributions are summarized as follows.
\begin{itemize}
  \item We propose, for the first time, a scheme for CFmMIMO networks to support any FL framework. In this scheme, any iterative algorithm can be developed to optimize the FL performance before the FL process is executed.
  %The design of this algorithm rely on specific optimization purposes (e.g., training time or loss function minimization).
  Each instead of all iterations of this FL optimization algorithm or the FL process happens within one $\widetilde{T}_{c}$ in order to guarantee channel stability during its operation.
  In each iteration of the FL process, we propose using the APs as relays to transmit the training updates between the CPU and UEs. Doing so allows any beamforming/filtering design to be applied to the APs in order to enhance the performance of training update transmission.
  %For ease of implementation, we apply a conjugate beamforming/matched filtering scheme to the APs, with the CSI acquired via uplink (UL) training pilots.

  %Effectively, the CPU and UEs are respectively the central server and mobile devices, while the APs relay the training updates between the CPU and UEs.

  %The steps of transmitting and computing training updates of the FL process are optimized for training time minimization.
%      Every step of FL algorithms
%%      , e.g, global updates at the central server, downlink (DL) training updates, local updates at the UEs, and UL training updates
%      is supported by each step in the proposed framework.
  \item To show how to optimize the FL performance using the proposed scheme, we consider an existing FL framework \cite{ma15OMS} as an example and target the key performance metric of ``training time minimization". We formulate a mixed-timescale stochastic nonconvex optimization problem that minimizes the time of one FL process.
      The formulated problem captures the complex interactions among the FL training time, and transmission and computation of FL training updates in a CFmMIMO network.
      Here, a conjugate beamforming/matched filtering scheme is applied to the APs for ease of implementation.
      The local accuracy, power control, data rate and UE's processing frequency are jointly designed, subject to the practical constraints on UEs' energy consumption and imperfect channel estimation.
  \item Utilizing the general framework in \cite{liu18TSP}, we propose a new algorithm
  %Each iteration solves a long-term optimization (LTO) problem to update the local accuracy and a short-term optimization (STO) subproblem to update the remaining variables.
  %To deal with the stochastic nature of the LTO problem, we replace its cost function by a convex surrogate function.
  %To deal with the nonconvex nature of the STO subproblem, we recast it as an approximated problem which is then solved by successive convex quadratic programming.
  that is proven to converge to at least the neighborhood of the stationary points
  %\footnote{The definition of stationary points is given in \cite[Definition 1]{liu18TSP}.}
  of the formulated problem. Here, the coupling among the variables makes it challenging to develop a specific algorithm that satisfies all the strict conditions stated in the general framework of \cite{liu18TSP}. It is also noted that our algorithm only requires channel stability in each but not all iterations. This important feature ensures the problem and its solution remain up-to-date and valid during the running time of the algorithm, despite the channel variations.
  %Note that \cite{liu18TSP} only proposes a general framework and provides theoretical conditions to design an algorithm for solving two-stage stochastic nonconvex optimization problems. Here, our contribution lies in designing a novel and specific algorithm satisfying all these conditions to solve the considered problem.
  \item Simulation results verify the convergence of the proposed algorithm, and show that our solution reduces the training time by up to $55\%$ compared with the baseline schemes.
  %They also demonstrate the importance of optimizing local accuracy and transmit power on reducing the training time.
  %In addition, the impacts of local accuracy, UE's processing frequency, UE's energy usage limit, and the length of UL training pilots on the training time are also revealed.
  They further confirm that CFmMIMO offers the lowest training time compared to cell-free TDMA massive MIMO and collocated massive MIMO.
%      They can achieve a much lower training time than the cell-free TDMA massive MIMO systems while significantly outperforming the collocated massive MIMO systems at large coverage.
\end{itemize}
%The APs use CSI (acquired via the uplink training) and simple conjugate beamforming/matched filter to serve all uplink and downlink users over the same time and frequency resources.
%We optimize the training speedup by finding the optimal solutions of power control, UE's CPU frequency, and local accuracy. The problem is constrained by the maximum energy consumed at each UE.
%In numerical results, we show that ...

\textit{Paper Organization and Notation:}
The rest of this paper is organized as follows. Section~\ref{sec:FL} proposes a novel scheme for a CFmMIMO network to support a general FL framework.
Section~\ref{sec:FLexample} introduces a specific example of the general FL framework considered in this paper.
Section~\ref{sec:SystemModel} presents the system model and assumptions. Section~\ref{sec:PF} formulates the FL training time minimization problem, whereas Section~\ref{sec:alg} proposes a new algorithm to solve the formulated problem.
For comparison, Section~\ref{sec:TDMACOL} introduces cell-free TDMA and collocated  massive MIMO systems to also support the considered FL framework.
Section~\ref{sec:sim} verifies the performance of the developed algorithm through comprehensive numerical examples. Finally, Section~\ref{sec:con} concludes the paper.

%\emph{Notation:}
In this paper, boldfaced symbols are used for vectors and capitalized boldfaced symbols for matrices.
$\pmb{X}^*$ and $\pmb{X}^H$ are the conjugate and conjugate transposition of a matrix $\pmb{X}$, respectively.
$\RRR^d$ is a space where its elements are real vectors with length $d$.
$\langle\x,\y\rangle$ means the inner product of vectors $\x$ and $\y$.
%$\pmb{I}$ and $\pmb{0}$ are the identity and zero matrices with appropriate dimensions, respectively.
%the real part of a complex number $x$ is denoted as $\mathfrak{R}\{x\}$.
%For a scalar $x$, $\lfloor x\rfloor$ denotes the largest integer that is not larger than $x$.
%$\pmb{X}$ and $\pmb{x}$ are respectively denoted as a complex matrix and a vector.
$||.||$  denotes the $\ell_2$-norm function.
%$||.||_0$ and $\one_{\{.\}}$ denotes the $\ell_0$-norm and the indicator function, respectively.
$\CN(\pmb{\mu},\pmb{Q})$ denotes the circularly symmetric complex Gaussian distribution with mean $\pmb{\mu}$ and covariance $\pmb{Q}$. $\nabla g$ is the gradient of a function $g$.
%$|\GG|$ stands for the number of elements in set $\GG$.
$\EEE\{x\}$ denotes the expected value of a random variable $x$.
\vspace{-3mm}

\section{Proposed Scheme for CFmMIMO Networks to Support FL}
\label{sec:FL}
\subsection{The General FL Framework}
A global ML problem is solved at a central server with
a global training data set partitioned over a number of participating clients. %which are mobile devices in wireless networks.
Each client trains their local model by an arbitrary learning algorithm. Let $\K=\{1,...,K\}$ be the set of clients and $D_k$ the size of the local data stored at client $k$. Then $\widetilde{D}=\sum_{k\in\K}D_k$ is the size of the global training data. Denote by $\DD=\{1,...,\widetilde{D}\}$ and $\DD_k=\{1,...,D_k\}$ the index sets of the global data samples and the local data samples at a client $k$, respectively. In a typical supervised learning, a data sample $i\in\DD$ is defined as an input-output pair $\{\x_i\in\mathbb{R}^d,y_i\}$.

For $\lambda>0$, the general global ML problem can be posed as the following minimization \cite{mcmahan17,ma15OMS}
\vspace{-1mm}
\begin{align}\label{FL:glob:prob}
\underset{\w\in\mathbb{R}^d}{\min} \,\,
J(\w) \triangleq \frac{1}{\widetilde{D}}\sum_{i\in\DD}f_i(\w) + \lambda g(\w),
\end{align}
\vspace{-3mm}

\noindent
where $f_i(\w)$ is the loss function at data sample $i$ and $g(\w)$ is a regularization term with a model parameter $\w$. Some popular examples are $f_i(\w) = \frac{1}{2}(\x_i^T\w-y_i)^2$ for a linear regression problem and $f_i(\w)=\{0,1-y_i\x_i^T\w\},y_i\in\{-1,1\}$ for a support vector machine. Here, the learning problem is to find $\w$ that characterizes the output $y_i$ with the loss function $f_i(\w)$ for a given input $\x_i$. Note that $f_i(\w)$ is not necessarily convex.

In a general FL framework to solve the general ML problem \eqref{FL:glob:prob}, this problem is decomposed into $K$ separate local ML problems that are solved at $K$ clients in parallel. For ease of presentation, we make the following definitions.
\vspace{-1mm}

\noindent
\begin{definition}
``Global DL training update'' is the information sent from the central server to the clients. Similarly, ``global UL training updates'' are those from the clients to the central server.
\end{definition}
\vspace{-1mm}

\noindent
The general FL framework is described in Algorithm~\ref{alg:FLgeneral}. Each iteration of Algorithm~\ref{alg:FLgeneral} consists the four key steps (S1)-(S4).
\begin{definition}
``An FL process'' is defined as a full execution of Algorithm~\ref{alg:FLgeneral}.
\end{definition}
\begin{algorithm}[!t]
\caption{A general FL framework}
\begin{algorithmic}[1]\label{alg:FLgeneral}
%\small{
\STATE \textbf{Input}: $n=1$, an initial global DL training update
\REPEAT
\STATE (S1) The central server sends the global DL training update to the UEs.
\FOR{$k\in\K$ in parallel}
\STATE (S2) Client $k$ updates and solves its local ML problem on its local data set and then computes the global UL training update
\STATE (S3) Client $k$ sends its computed global UL training update to the central server
\ENDFOR
\STATE (S4) The central server computes the global DL training update by aggregating the received UL training updates.
\STATE Update $n=n+1$
\UNTIL{convergence with the global accuracy $\epsilon$}
%}
\end{algorithmic}
\end{algorithm}
\vspace{-2mm}

\noindent
\begin{remark}
The designs of local ML problems at the clients, the global DL/UL training updates, and the types of aggregation of training updates at the central server are different according to the different designs of FL frameworks for different types of objective functions due to different ML applications \cite{ahn19arXiv,mohri19arXiv,hangyu18arXiv,konecny17OMS,ma15OMS,jaggi14NIPS}.
\end{remark}
\vspace{-4mm}

\noindent
\subsection{Proposed Cell-Free Massive MIMO Network Structure to Support the General FL Framework}
\vspace{-1mm}

\noindent
\label{subsec:CFmMIMO}
Here, we propose using the CFmMIMO network structure \cite{ngo17TWC} illustrated in Fig.~\ref{fig:systemmodel} to support the general FL framework discussed above.
In this structure, a central processing unit (CPU) is connected to a set of access points (APs) $\MM=\{1,...,M\}$ via backhaul links with sufficient capacities. %\footnote{In practice, the backhaul link capacity will be subject to practical constrains \cite{bashar19TCOM,bashar19TGCN}. The study on the impact of limited-capacity backhaul links are needed and left for future works.}.
These APs serve a given set $\K$ of participating UEs via wireless access links at the same time and in the same frequency bands\footnote{In general, since the UEs are geographically distributed, some user may loss the connectivity, and hence, cannot participate in the FL process. However, one of the main strong property of CFmMIMO is that, with very high probability, it can provide uniformly good service for all users in the networks \cite{ngo17TWC}. In the other words, in CFmMIMO, the connectivity probability of each user is very high. Therefore, in the paper, for simplicity, we assume that all UEs participate in the FL process.}. The APs and UEs are each equipped with a single antenna.
%This system model is similar to that in \cite{ngo17TWC}.
%In this network,
The CPU and UEs act as the central server and the clients in the general FL framework, respectively. The APs are used to relay the training updates between the CPU and the UEs %Furthermore, we insist a functional platform is installed at the CPU to interact with the UEs via an application interface, in order to leverage FL and build a global model of ML.
\vspace{-1mm}

\noindent
\begin{remark}
Both the AP and UEs can be considered as the clients in the general FL framework. In this paper, we choose the UEs to be the clients instead of the APs \cite{tran19INFOCOM}. By doing so, data privacy is more protected since raw data does not have to be shared over wireless networks. Moreover, the computational resource of the UEs can be empowered with new computing engines such as Snapdragon $835$ \cite{qualcomm17}. It is possible to push the ML tasks further to the UEs \cite{ti19arXiv,tran19INFOCOM}.
\end{remark}
\vspace{-5mm}

\noindent
\subsection{Proposed Scheme for a CFmMIMO to Support the General FL framework}
\label{subsec:proscheme}
\begin{figure*}[t!]
\centering
\includegraphics[width=0.61\textwidth]{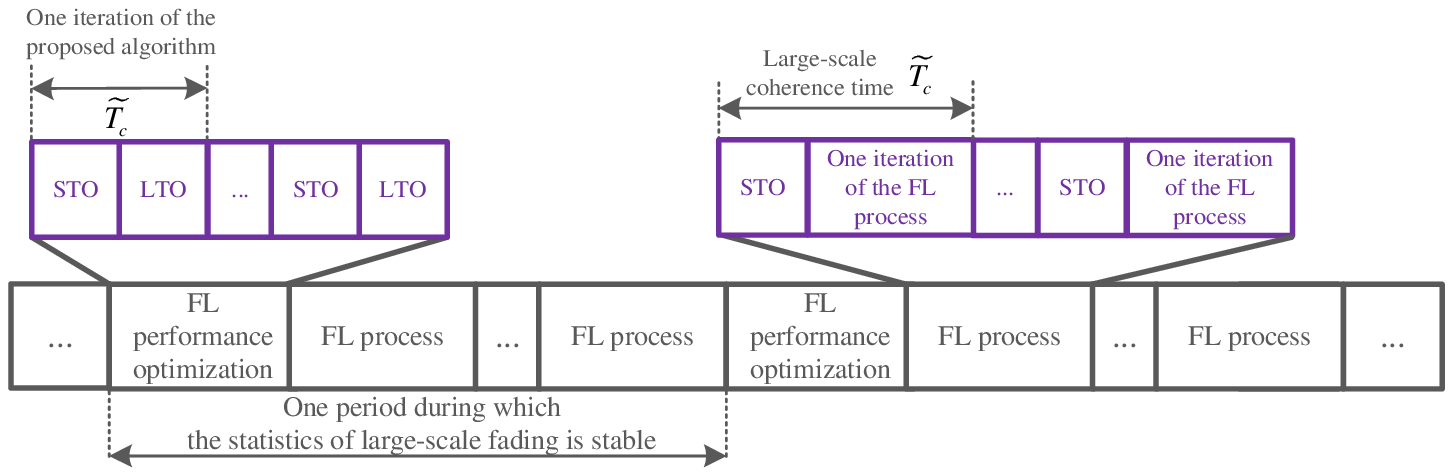}
\caption{Proposed scheme of a CFmMIMO system to support FL}
\label{fig:time1}
\centering
\includegraphics[width=0.56\textwidth]{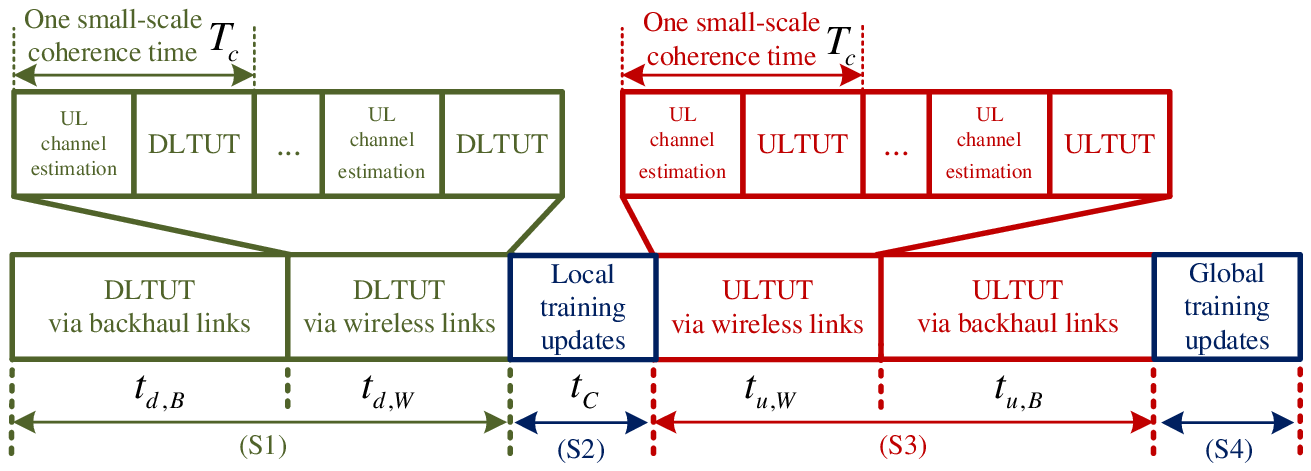}
\caption{One iteration of the FL process}
\label{fig:time2}
\end{figure*}

Now, to realize an FL process in the considered CFmMIMO network structure, we propose a general scheme shown in Fig.~\ref{fig:time1}. As seen from the figure, we divide the time period during which the statistics of large-scale fading is stable into multiple time intervals. The first interval of ``FL performance optimization" is used for optimizing the performance of FL.
%In this ``FL performance optimization" interval, any algorithms can be developed to pre-optimize the performance of the FL process.
The remaining intervals are reserved for the FL process; hence, termed as ``FL process" intervals.

\subsubsection{Optimizing the performance of FL}
%Let us now focus on the ``FL performance optimization" interval.
In the ``FL performance optimization" interval, any algorithms can be developed to optimize the performance of FL before the FL processes are executed.
Here, we denote by ``system parameters" the parameters that are designed by the FL optimization algorithm. These parameters are grouped into short-term and long-term parameters. The short-term parameters change in the timescale of large-scale coherence times, whereas the long-term parameters the statistics of large-scale fading.
In each iteration of the optimization algorithm, the short-term parameters are designed in the short-term optimization (STO) time blocks, whereas the long-term parameters in the long-term optimization (LTO) time blocks.
As a result of the channel hardening effect in massive MIMO, the wireless channel remains unchanged during one large-scale coherence time $\widetilde{T}_c$.
In practice, the completion time of the optimization algorithm can be larger than $\widetilde{T}_{c}$.
%Optimizing all system parameters in one $\widetilde{T}_{c}$ would make the optimization problem easily obsolete before the algorithm terminates.
Therefore, we insist that only one iteration of the algorithm is to happen within $\widetilde{T}_{c}$.
%Each pair of STO-LTO is computed again when the large-fading coefficients changes.
%The number of STO-LTO pairs is the number of iterations for the algorithm to converge.

\subsubsection{Implementing the FL process}
%Next, we concentrate on the ``FL process intervals'' in which the FL processes are executed.
In the ``FL process'' intervals, the FL process is executed with the long-term parameters obtained from the ``FL performance optimization'' interval.
%remain unchanged until there is any change in the statistics of large-scale fading.
The short-term parameters are optimized in the STO time block to enhance the performance of training update transmission  before each iteration of the FL process.
Since the completion time of the FL process can be larger than $\widetilde{T}_{c}$,
we insist that both STO and ``one iteration of the FL process'' time blocks are to happen in one $\widetilde{T}_{c}$.
Here, we note that the results of the LTO time blocks remain unchanged for several FL process, while the results of the STO time blocks are invariant only in one iteration of an FL process. Therefore, after the ``FL performance optimization'' step is executed, only the results of the LTO time block are used in FL processes. In each iteration of an FL process, the results of the STO time block are not the results of the STO time block in the ``FL performance optimization'' step, but rather are computed by the same method used for the STO time block in the ``FL performance optimization'' step.

%The number of pairs of STO and one iteration of the FL process is the number of iterations $G(\theta)$ for one FL process to terminate.
\subsubsection{Implementing each iteration of the FL process}
Fig.~\ref{fig:time2} shows the time block of ``one iteration of the FL process'', in which the intervals of the four key steps (S1)-(S4) to implement each iteration of the FL process by the CFmMIMO network model are illustrated.
%It includes the intervals for global DL training update, local training update, global UL training update, and global UL training update.
Here, the interval of Step (S1) in Algorithm~\ref{alg:FLgeneral} consists one interval of ``DL training update transmission (DLTUT) via backhaul links'' from the CPU to the APs, and one interval of ``DLTUT via wireless links'' from the APs to the UEs.
The interval of Step (S3) includes one interval of ``UL training udpate transmission (ULTUT) via wireless links'' from the UEs to the APs, and one interval of ``ULTUT via backhaul links'' from the APs to the CPU.

As also seen from Fig.~\ref{fig:time2}, we split each time block of ``DLTUT via wireless links'' or ``ULTUT via wireless links'' into multiple intervals. The intervals of ``UL training'' is used for channel estimation. The remaining intervals are used for DL/UL training update transmission\footnote{The proposed scheme above focuses only on support FL. Therefore, in each small-scale coherence time, all the times not used to estimate the channel are used for transmitting training updates. The scheme that supports FL and data transmission at the same time is left for future works.}. In these ``DLTUT'' or ``ULTUT'' intervals, any beamforming/filtering design can be applied to optimize the performance of training update transmission.
Here, we insist that the pair of ``UL training'' and ``DLTUT''/``ULTUT'' intervals happen in one small-scale coherence time $T_c$ in order to adapt to the variation of small-scale fading\footnote{In this work, we consider time-division duplexing (TDD) where channels are first estimated at the APs via uplink channel estimation. Owing to the channel reciprocity property in massive MIMO, these channel estimates will then be used for: (i) precoding the data symbols in the downlink data transmission, and (ii) used for combining the received signals in the uplink data transmission. Moreover, there is no need for downlink channel estimation because of the channel hardening property in massive MIMO systems. This transmission protocol is widely used in the massive MIMO literature (see, e.g., \cite{emil17}).}.
In addition, the data size of training updates may be larger than the amount of data that can be transmitted in $T_c$. As such, there may be one or several ``DLTUT'' or ``ULTUT'' intervals in each ``DLTUT via wireless links'' or ``ULTUT via wireless links'' time block in order to complete the transmission of one global DL/UL training update.
\vspace{-1mm}

\noindent
\begin{remark}
The proposed CFmMIMO network model in Section~\ref{subsec:CFmMIMO} and the proposed scheme in Section~\ref{subsec:proscheme} can be used to support any version of the general FL framework to solve any global ML problem. This scheme only requires developing specific algorithms to optimize the performance of specific FL frameworks. In the following, we consider a specific example of the general FL framework and investigate an algorithm to optimize the FL performance of this framework.
\end{remark}
\vspace{-3mm}

\noindent
\section{A Specific Example of the General FL Framework}
\label{sec:FLexample}
\vspace{-1mm}

\noindent
Given the proposed framework in Section II, this section considers a specific example of the general FL framework to show how the performance of an FL process can be optimized in the latter sections. In particular, the FL optimization problem for this example is introduced in Section~\ref{sec:PF}, and the algorithms to solve this problem are proposed in Section~\ref{sec:alg}.

Because of all the potential advantages offered by FL, many versions \cite{ahn19arXiv,mohri19arXiv,hangyu18arXiv,konecny17OMS,ma15OMS,jaggi14NIPS} of the general FL framework have so far been studied despite research on FL is still in its infancy. Here, we consider an existing FL framework of \cite{ma15OMS} which is briefly introduced as follows.
%This FL framework is briefly introduced in the following.}
In this FL framework, the considered loss function $f_i$ is convex and the dual problem of \eqref{FL:glob:prob} is written as
\vspace{-2mm}
\begin{align}\label{FL:glob:prob:dual}
\underset{\ALPHA\in\mathbb{R}^{\widetilde{D}}}{\max} \,\,
J_{\widetilde{D}}(\ALPHA) \triangleq \tfrac{1}{\widetilde{D}}\sum_{i\in\DD}-f_i^*(-\ALPHA) - \lambda g^*(\CHI(\ALPHA)),
\end{align}
\vspace{-4mm}

\noindent
which is a special case of the Fenchel duality \cite{ma15OMS}, where $f^*_i$ and $g^*$ are the convex conjugate functions of $f_i$ and $g$, respectively; $\ALPHA$ is a dual variable; $\CHI(\ALPHA)=\tfrac{1}{\lambda \widetilde{D}}\X\ALPHA$; $\X=[\x_1,...,\x_{\widetilde{D}}]\in\mathbb{R}^{d\times \widetilde{D}}$. It can be shown that if $\ALPHA^*$ is an optimal solution of \eqref{FL:glob:prob:dual}, then $\w(\ALPHA^*)=\nabla g^*\big(\tfrac{1}{\lambda \widetilde{D}}\X\ALPHA^*\big)$ is the optimal solution of \eqref{FL:glob:prob}. This property allows handling the dual variable $\ALPHA\in\mathbb{R}^{\widetilde{D}}$ instead of $\w\in\mathbb{R}^d$.
Since each $\alpha_i$ corresponds to each data sample $i$, $\ALPHA$ can be distributed in the same way that data are partitioned for the $K$ clients.

%We now define a new notation that is dependent on this partitioning as follows\footnote{Duy: revise - sounds unnatural!}.
Let $(\pphi)_i$ be the $i$-th element of vector $\pphi$ and $(\X)_i$ be the $i$-th column vector of matrix $\X$.
For any $\pphi\in\mathbb{R}^{\widetilde{D}}$, denote by $\pphi_{[k]}\in\mathbb{R}^{\widetilde{D}}$ the vector for device $k$, i.e., $(\pphi_{[k]})_i=(\pphi)_i$ if $i\in\mathcal{D} _k$ and $0$ otherwise. Similarly, for any $\X\in\mathbb{R}^{d\times \widetilde{D}}$,
denote by $\X_{[k]}\in\mathbb{R}^{d\times \widetilde{D}}$ the matrix for device $k$, i.e., $(\X_{[k]})_i=(\X)_i$ if $i\in\mathcal{D} _k$ if $i\in\DD_k$ and $\pmb{0}$ otherwise.  The local problem at each client $k$ is then assigned to find the optimal change $\pphi_{[k]}$ in the local dual variable $\ALPHA_{[k]}$, for a given previous $\ALPHA$ as
\vspace{-1mm}
 \begin{align}\label{FL:loc:prob}
\underset{\pphi_{[k]}\in\RRR^{\widetilde{D}}}{\max} \,\,
J_k(\pphi_{[k]},\CHI(\ALPHA),\ALPHA_{[k]}),
\end{align}
\vspace{-5mm}

\noindent
where
\vspace{-2mm}
\begin{align}\label{}
\nonumber
&J_k(\pphi_{[k]},\CHI(\ALPHA),\ALPHA_{[k]})
\\
\nonumber
&=-\tfrac{1}{K}g^*(\CHI(\ALPHA))
-\left\langle \tfrac{1}{\widetilde{D}}\X_{[k]}^T\nabla g^*(\CHI(\ALPHA)),\pphi_{[k]}\right\rangle
\\
& -\tfrac{\lambda}{2}\left\| \tfrac{1}{\lambda \widetilde{D}}\X_{[k]}\pphi_{[k]}\right\|^2
-\tfrac{1}{\widetilde{D}}\sum_{i\in\D_k}f_i^*(-(\ALPHA_{[k]})_i-(\pphi_{[k]})_i)
\end{align}
\vspace{-3mm}

\noindent
is the quadratic approximation of $J_{\widetilde{D}}$ at the dual variable $(\ALPHA_{[k]}+\pphi_{[k]})$. Here, the only information shared between the clients and the central server is the change in $\CHI$.

At iteration $n$, each client $k\in\K$ solves \eqref{FL:loc:prob} by an arbitrary iterative algorithm with a local accuracy level of $\theta$ in order to obtain its optimal solution $\pphi_{[k]}^*$. The local dual variable is updated as
\vspace{-2mm}
\begin{align}\label{update:dual}
\ALPHA_{[k]}^{(n+1)}=\ALPHA_{[k]}^{(n)} + \pphi_{[k]}^*.
\end{align}
\vspace{-5mm}

\noindent
Each client $k\in\K$ then shares its local change in $\CHI^{(n)}$, i.e.,
\vspace{-1mm}
\begin{align}\label{update:deltachi}
\Delta\CHI_k^{(n)}=\tfrac{1}{\lambda \widetilde{D}}\X_{[k]}\pphi_{[k]}^*,
\end{align}
\vspace{-5mm}

\noindent
to the central server. The central server aggregates the local information $\Delta\CHI_k^{(n)}$ received from all the clients and updates $\CHI$ as
\vspace{-2mm}
\begin{align}\label{update:chi}
\CHI^{(n+1)}=\CHI^{(n)} + \tfrac{1}{K}\Delta\CHI_k^{(n)}.
\end{align}
\vspace{-5mm}

\noindent
$\CHI^{(n+1)}$ is finally sent back to the clients to solve \eqref{FL:loc:prob}.
This process will terminate when a global accuracy level of $\epsilon$ is reached.
The FL framework described above is summarized in Algorithm~\ref{alg:FLexample}. Each iteration of Algorithm~\ref{alg:FLexample} also consists of four key steps (S1)-(S4) as that of Algorithm~\ref{alg:FLgeneral}.
\begin{algorithm}[!t]
\caption{FL framework \cite{ma15OMS}}
\begin{algorithmic}[1]\label{alg:FLexample}
%\small{
\STATE \textbf{Input}: $n=1$, an initial point $\ALPHA^{(0)}$ and an initial global DL training update $\CHI^{(0)}=\frac{1}{\lambda \widetilde{D}}\X\ALPHA^{(0)}$
\REPEAT
\STATE (S1) The CPU sends $\CHI^{(n)}$ to the UEs
\FOR{$k\in\K$ in parallel}
\STATE (S2) UE $k$ solves \eqref{FL:loc:prob} by an iterative algorithm with a local accuracy $\theta$ to obtain an optimal solution $\pphi_{[k]}^*$, and then updates
%the local dual variable $\ALPHA_{[k]}^{(n+1)}$ by \eqref{update:dual} and
the global UL training update $\Delta\CHI_k^{(n)}$ by \eqref{update:deltachi}
\STATE (S3) UE $k$ sends $\Delta\CHI_k^{(n)}$ to the CPU
\ENDFOR
\STATE (S4) The CPU updates $\CHI^{(n+1)}$ as \eqref{update:chi} and sends it back to the UEs
\STATE Update $n=n+1$
\UNTIL{convergence with the global accuracy $\epsilon$}
%}
\end{algorithmic}
\end{algorithm}

We assume that each client $k\in\K$ uses optimization algorithms such as stochastic average gradient (SAG) and stochastic variance reduced gradient (SVRG) to solve \eqref{FL:loc:prob} with a local accuracy level of $\theta$. The number of local iterations  is then \cite{konecny17OMS}
\vspace{-2mm}
\begin{align}\label{L}
L(\theta) = \nu\log(\tfrac{1}{\theta}),
\end{align}
\vspace{-5mm}

\noindent
where $\nu>0$ depends on the data size and structure of the local problem \cite{konecny17OMS}.
On the other hand, for strongly convex objective functions and the global accuracy level of $\epsilon$, %characterized as $\OO\left(\log(\frac{1}{\epsilon})\right)$
the number of global iterations are given by \cite{ma15OMS}
\vspace{-1mm}
\begin{align}\label{G}
  G(\theta) = \frac{\vartheta\log\left(\frac{1}{\epsilon}\right)}{1-\theta},
\end{align}
\vspace{-5mm}

\noindent
where
$\vartheta>0$ is a factor that depends on the characteristic and size of the whole data set \cite[Theorem 4.2]{ma15OMS}. Here, we assume that the characteristic and size of the whole data set does not change over the FL processes. Therefore, $\vartheta$ is constant and known \cite{tran19INFOCOM}.
\begin{remark}
In more complex ML models such as deep neural networks, the loss function $f_i$ is usually non-convex. Therefore, instead of the framework in \cite{ma15OMS}, more advanced FL frameworks are needed. In this paper, we choose to consider the FL framework in \cite{ma15OMS} for simpler ML models because it provides a clear relationship \eqref{G} between the number of global training updates and the local accuracy for ease of optimizing FL performance (see the next sections). Such a clear relationship is hard to find in the advanced FL frameworks with non-convex loss functions.
\end{remark}
\vspace{-4mm}

\noindent
\section{Detailed System Model to Support FL}
\label{sec:SystemModel}
This section details the CFmMIMO system model used to support the transmission and computation of the training updates in each iteration of the FL process (see Fig.~\ref{fig:time2}).
\vspace{-5mm}

\noindent
\subsection{Steps (S1) and (S3) in Each Iteration of the FL Process: Model of Training Update Transmission}
\vspace{-1mm}

\noindent
\subsubsection{UL channel estimation}
Denote by $\tau_c=T_cB_c$ the number of samples of each coherence block, where $B_c$ the coherence bandwidth.
UL pilot sequences are sent by all the UEs to all the APs simultaneously.
Denote by $\tau_{t}$ (samples) the length of one pilot sequence.
Let $\sqrt{\tau_{t}}\VARPHI_{k}\in\C^{\tau_{t}\times 1}$ be the pilot sequence transmitted from UE $k\in\K$, where $\|\VARPHI_{k}\|^2=1,\forall k\in\K$.
The channel from a UE $k$ to an AP $m$
is modeled as $g_{mk} = (\beta_{mk})^{1/2}\tilde{g}_{mk}$,
where $\beta_{mk}$ and $\tilde{g}_{mk}\in\C$ represent the large-scale fading and small-scale fading channel coefficients, respectively. Assume that $\tilde{g}_{mk}$ is an independent and identically distributed (i.i.d.) $\CN(0,1)$ random variable.

The AP $m$ receives the pilot vector $\y_{m} = \sqrt{\tau_{t}\rho_{t}}\sum_{k\in\K}g_{mk}\VARPHI_{k} + \w_{m}$,
where $\rho_{t}$ is the normalized signal-to-noise ratio (SNR) of each pilot symbol, and $\w_m\in\CN(\pmb{0},\pmb{I})$ is the additive noise at the AP $m$.
The projection of $\y_{m}$ onto $\VARPHI_{k}$ is given as $\hat{y}_{mk}
=\VARPHI_{k}^H\y_{m} =
\sqrt{\tau_{t}\rho_{t}}\sum_{\ell\in\K}g_{m\ell}\VARPHI_{k}^H\VARPHI_{\ell} + \VARPHI_{k}^H\w_{m}$.
After receiving $\hat{y}_{mk}$, the AP $m$ estimates $g_{mk}$ by using the minimum mean-square error (MMSE) estimation. Given $\hat{y}_{mk}$, the MMSE estimate $\hat{g}_{mk}$ of $g_{mk}$ is obtained as \cite{kay93}: $\hat{g}_{mk}
= \EEE\{\hat{y}_{m}^*g_{mk}\}(\EEE\{|\hat{y}_{mk}|^2\})^{-1}\hat{y}_{mk}
= c_{mk}\hat{y}_{mk}$,
where $c_{mk}\triangleq\frac{\sqrt{\tau_{t}\rho_{t}}\beta_{mk}}
{\sum_{\ell\in\K}\tau_{t}\rho_{t}\beta_{m\ell}|\VARPHI_k^H\VARPHI_\ell|^2+1}$.
From the property of MMSE channel estimation, $\hat{g}_{mk}$ is distributed according to $\CN(0,\sigma_{mk}^2)$, where
$\sigma_{mk}^2=\frac{\tau_{t}\rho_{t}(\beta_{mk})^2}
{\sum_{\ell\in\K}\tau_{t}\rho_{t}\beta_{m\ell}|\VARPHI_k^H\VARPHI_\ell|^2+1}$ \cite{kay93}.
\begin{remark}
In indoor communications, the time for UL training can be much smaller than the small-scale coherence time $T_c$. For example, a system supporting users' mobility of $v=0.75~\text{m/s}=2.7~\text{km/h}$, delay spread of $T_d=0.5~\mu\text{s}$ and carrier frequency $f_c=2~\text{GHz}$ has a small-scale coherence time of $T_c=\frac{c}{4f_cv}=50~\text{ms}$, and coherence bandwidth $B_c=\frac{1}{2T_d}=1~\text{MHz}$  \cite{emil17}. Suppose $\tau_t=20$, the time for UL channel estimation is $t_{ce} = \frac{\tau_t}{B_c}=0.02~\text{ms}\ll T_c$. Therefore, the time for UL channel estimation can be ignored in one small-scale coherence time and one FL process interval.
\end{remark}
\vspace{-1mm}

\noindent
\subsubsection{Step (S1) in each iteration of the FL process}
At the CPU, the global DL training update intended for a UE $k$ is
encoded into a symbol $s_{d,k} \sim \CN(0,1)$. Here, each DL/UL training update is considered as a data message that is widely used in the literature of wireless communications \cite{ngo16}. The CPU then sends $s_{d,k}, \forall k\in\K$, to all the APs over backhaul links.
Let $S_d$ (bits) and $R_{d,k}$ (bps) be the data size and the data rate of the global DL training update for the UE $k$, respectively. The download latency from the CPU to all the APs is given by
\vspace{-3mm}
\begin{align}\label{latency:d:bachhaul}
t_{d,B}(\RR_d) = \frac{KS_d}{\sum_{k\in\K}R_{d,k}},
\end{align}
\vspace{-4mm}

\noindent
where $\RR_d\triangleq\{R_{d,k}\}_{k\in\K}$.

For ease of implementation, we apply a conjugate beamforming scheme to the APs to precode the message signals before wirelessly transmitting them to the UEs (using the channel estimates from the UL channel estimation). The transmitted signal at an AP $m$ is expressed as $x_{d,m} = \sqrt{\rho_{d}}\sum_{k\in\K}\sqrt{\eta_{mk}}(\hat{g}_{mk})^*s_{d,k}$,
where $\rho_{d}$ is the maximum normalized transmit power (normalized by the noise power $N_0$) at each AP
and $\eta_{mk}, \forall m\in\MM,k\in\K$, is a power control coefficient.
%$\hat{\g}_{mk}^d$ is the column $k$ of $\hat{\g}_{d,m}$,
%and $\s^d\triangleq[s_1^d;...;s_{K_d}^d]\in\C^{K_d\times 1}$.
The AP $m$ is required to meet the average normalized power constraint, i.e., $\EEE\{|x_{d,m}|^2\}\leq \rho_d$, which can also be expressed as the following per-AP power constraint:
\vspace{-1mm}
\begin{align}\label{power:d:cons}
\sum_{k\in\K}\sigma_{mk}^2\eta_{mk} \leq 1, \forall m.
%\sum_{k\in\K_d}\eta_{mk}\sigma_{d,mk}^2 \leq 1, \forall m.
\end{align}
\vspace{-4mm}

\noindent

The received signal at the UE $k$ is given by $r_{d,k} =
\sum_{m\in\MM} g_{mk}x_{d,m} + w_{k}$,
where $w_{k}$ is the additive noise $\CN(0,1)$ at the UE $k$.
The achievable DL rate at the UE $k$ is
\vspace{-5mm}
\begin{align}\label{rate:d}
R_{d,k} \leq  h_{d,k}(\ETA),
\end{align}
\vspace{-6mm}

\noindent
where $\ETA\triangleq\{\eta_{mk}\}_{m\in\MM,k\in\K}$ and $h_{d,k}(\ETA)$ is given in \eqref{h:d} shown at the top of the page \cite{ngo17TWC}.
\begin{figure*}[t!]
\begin{align}\label{h:d}
 h_{d,k}(\ETA) =\frac{\tau_c-\tau_t}{\tau_c}B\log_2\Bigg(1+\frac
{\rho_d\left(\sum_{m\in\MM}\eta_{mk}^{1/2}\sigma_{mk}^2\right)^2}
{\rho_d\sum_{\ell\in\K\setminus k}
\left(\sum_{m\in\MM}\eta_{m\ell}^{1/2}\sigma_{m\ell}^2\frac{\beta_{mk}}{\beta_{m\ell}}\right)^2
|\VARPHI_\ell^H\VARPHI_k|^2
+\rho_d\sum_{\ell\in\K}\sum_{m\in\MM}\eta_{m\ell}\sigma_{m\ell}^2\beta_{mk}+1}\Bigg)
\end{align}
\hrulefill
\end{figure*}
Note that in \eqref{h:d}, $B$ is the bandwidth.
The download latency from the APs to the UE $k$ is given by
\vspace{-2mm}
\begin{align}\label{latency:d:wireless}
t_{d,k}(R_{d,k}) = \frac{S_d}{R_{d,k}}.
\end{align}
\vspace{-4mm}

\noindent
\subsubsection{Step (S3) in each iteration of the FL process}
After updating the local model, the UE $k$ encodes the global UL training update
into a symbol $s_{u,k} \sim \CN(0,1)$.
The symbol $s_{u,k}$ is then allocated a transmit amplitude value $\sqrt{\rho_{u}\zeta_k}$ to generate a baseband signal $x_{u,k}$ for wireless transmissions, i.e., $x_{u,k}=\sqrt{\rho_{u}\zeta_k}s_{u,k}$. The UE $k$ is adhered to the average transmit power constraint, i.e., $\EEE\left\{|x_{u,k}|^2\right\}\leq \rho_u$, which can also be expressed in a per-UE constraint as
\vspace{-2mm}
\begin{align}\label{power:u:cons}
0\leq\zeta_k\leq 1,\forall k\in\K.
\end{align}
\vspace{-6mm}

\noindent
The upload latency from the UE $k$ to the AP $m$ is given by
\vspace{-1mm}
\begin{align}\label{latency:u:wireless}
t_{u,k}(R_{u,k}) = \frac{S_u}{R_{u,k}},
\end{align}
\vspace{-4mm}

\noindent
where $S_u$ (bits) and $R_{u,k}$ (bps) are the data size and the data rate of the global UL training update, respectively.

The received signal at the AP $m$ is expressed as
\vspace{-2mm}
\begin{align}\label{signal:UP:receive}
\nonumber
y_{u,m}
&=
\sum_{k\in\K}g_{mk} x_{u,k} + w_{u,m}
\\
&=
\sqrt{\rho_u}\sum_{k\in\K}g_{mk}\sqrt{\zeta_{k}}s_{u,k}
+ w_{u,m},
\end{align}
\vspace{-4mm}

\noindent
where $w_{u,m}\sim\CN(0,1)$ is the additive noise.
To detect the message symbol transmitted from the UE $k$, the AP $m$ computes and sends $(\hat{g}_{mk})^*y_{u,m}$ to the CPU. The upload latency from the APs to the CPU is expressed as
\vspace{-1mm}
\begin{align}\label{latency:u:backhaul}
t_{u,B}(\RR_u) = \frac{KS_u}{\sum_{k\in\K}R_{u,k}},
\end{align}
\vspace{-5mm}

\noindent
where $\RR_u\triangleq\{R_{u,k}\}_{k\in\K}$.
%Similar to the DL transmission, the UL data transmission rates via the backhaul links are constrained by
%\begin{align}\label{cons:u:cap}
%\sum_{k\in\K}R_{u,k} \leq C_{u,\max}.
%\end{align}

At the CPU, the symbol $s_{u,\ell}$ is detected from the received signal $r_{u,k}$:
\vspace{-3mm}
\begin{align}\label{signal:CPU:receive}
\nonumber
r_{u,k} =
%&\underbrace{\sqrt{\rho_u}\sum_{m\in\MM}\sqrt{\zeta_{k}}(\hat{g}_{mk})^*
%g_{mk} s_{u,k}}_{\text{DS}_{u,k}}
&\sqrt{\rho_u}\sum_{m\in\MM}\sqrt{\zeta_{k}}(\hat{g}_{mk})^*
g_{mk} s_{u,k}
\\
\nonumber
%& + \underbrace{\sqrt{\rho_u}\sum_{m\in\MM}\sum_{\ell\in\K\setminus k}
%\sqrt{\zeta_{\ell}}(\hat{g}_{mk})^*g_{m\ell} s_{u,\ell}}_{\text{MUI}_{u,k}}
& + \sqrt{\rho_u}\sum_{m\in\MM}\sum_{\ell\in\K\setminus k}
\sqrt{\zeta_{\ell}}(\hat{g}_{mk})^*g_{m\ell} s_{u,\ell}
%&+ \underbrace{\sum_{m\in\MM}(\hat{g}_{mk})^*w_{u,m}}_{\text{N}_{u,k}},
\\
& + \sum_{m\in\MM}(\hat{g}_{mk})^*w_{u,m}.
\end{align}
\vspace{-4mm}

\noindent
The achievable UL rate for the UE $k$ is given by
\vspace{-1mm}
\begin{align}\label{rate:u}
R_{u,k} \leq h_{u,k}(\ZETA),
\end{align}
\vspace{-5mm}

\noindent
where $\ZETA\triangleq\{\zeta_{k}\}_{k\in\K}$ and $h_{u,k}(\ZETA)$ is defined in \eqref{h:u} shown at the top of the next page \cite{ngo17TWC}.
\begin{figure*}[t!]
\begin{align}\label{h:u}
h_{u,k}(\ZETA)= \frac{\tau_c-\tau_t}{\tau_c}B\log_2\Bigg(1+
\frac
{\rho_u\zeta_{k}\left(\sum_{m\in\MM}\sigma_{mk}^2\right)^2}
{\rho_u\sum_{\ell\in\K\setminus k}\zeta_{\ell}
\left(\sum_{m\in\MM}\sigma_{mk}^2\frac{\beta_{m\ell}}{\beta_{mk}}\right)^2
|\VARPHI_k^H\VARPHI_\ell|^2
+\rho_u\sum_{\ell\in\K}\zeta_{\ell}\sum_{m\in\MM}\sigma_{mk}^2\beta_{m\ell}
+\sum_{m\in\MM}\sigma_{mk}^2}
\Bigg)
\end{align}
\hrulefill
\end{figure*}
\vspace{-4mm}

\noindent
\subsection{Step (S2) in Each Iteration of the FL Process: Model of Local Training Update Computation at UEs}
\label{sec:SystemModel:B}
Denote by $c_k$ (cycles/sample) the number of processing cycles for a UE $k$ to process one data sample. $c_k$ is known \emph{a priori} by an offline measurement \cite{miettinen10}. Let $D_k$ (samples) and $f_k$ (cycles/s) be the size of the local data set and the processing frequency of the UE $k$, respectively.
The latency of computing the local training update at the UE $k$ is by
\vspace{-1mm}
\begin{align}\label{}
t_{C,k}(\theta,f_k) = L(\theta)\frac{D_kc_k}{f_k},
\end{align}
\vspace{-4mm}

\noindent
where $L(\theta)$ is the number of local training iterations (see \eqref{L}) and $\frac{D_kc_k}{f_k}$ is the time taken to compute the local update over its local training data set in each iteration.
Given the limited computational resource at the UEs, we only focus on the delay of computing the local updates at the UEs.
Since the computational resource of the CPU is much more abundant than that of the UEs, the latency of aggregating the global UL training updates at the CPU is negligible, and hence ignored.
\vspace{-5mm}

\noindent
\subsection{The Model of UE's Energy Consumption}
Because the time for UL channel estimation is negligible compared with  one  FL  training interval, the energy consumed in the time block of UL channel estimation is ignored.
The energy consumption in the time block of ULTUT at a UE $k$ is given by
\vspace{-2mm}
\begin{align}\label{}
  E_{T,k}(\zeta_k,R_{u,k}) = \rho_u N_0\zeta_{k}\frac{S_u}{R_{u,k}},
\end{align}
\vspace{-4mm}

\noindent
where $\zeta_{k}$ is the transmitted UL power and $\frac{S_u}{R_{u,k}}$ is the delay incurred by transmitting the global UL training update $\Delta\CHI_k$. The energy required for computing local training updates at the UE $k$ is expressed as \cite{tran19INFOCOM}
\begin{align}\label{}
  E_{C,k}(\theta,f_k) = L(\theta)\frac{\alpha}{2}c_kD_kf_k^2,
\end{align}
where $\frac{\alpha}{2}$ is the effective capacitance coefficient of the UEs' computing chipset.

\section{FL Training Time Minimization: Problem Formulation}
\label{sec:PF}
To optimize the performance of the FL process using the considered FL framework \cite{ma15OMS} in the CFmMIMO network model discussed in Sections~\ref{sec:SystemModel}, this paper targets the key performance metric of \emph{training time minimization}.

In each iteration of the FL process, the time of Step (S1) for a UE $k$ involves the transmission delay of sending the global DL training update from the CPU to the APs via backhaul links and that from the APs to UE $k$ via wireless links, i.e.,
\vspace{-1mm}
\begin{align}\label{}
t_{T,k}^d(\RR_d) = t_{d,B}(\RR_d) + t_{d,k}(R_{d,k}) =
\frac{KS_d}{\sum_{k\in\K}R_{d,k}}+\frac{S_d}{R_{d,k}}.
\end{align}
\vspace{-4mm}

\noindent
Similarly, the time of Step (S2) for the UE $k$ consists of the delay of transmitting the global UL training update from it to the APs and from the APs to the CPU, i.e.
\vspace{-1mm}
\begin{align}\label{}
%\nonumber
t_{T,k}^u (\RR_u)
=  t_{u,k}(R_{u,k}) + t_{u,B}(\RR_u)
= \frac{S_u}{R_{u,k}}+\frac{KS_u}{\sum_{k\in\K}R_{u,k}}.
\end{align}
\vspace{-4mm}

\noindent
In the proposed scheme in Section~\ref{subsec:proscheme}, each of the steps (S1)-(S4) of one iteration of the FL process must be completed for all the UEs before the latter step is executed. Therefore, the time of one iteration of the FL process is
\vspace{-1mm}
\begin{align}\label{totaltime:global}
\nonumber
&T_G(\theta,\f,\RR_d,\RR_u)
\\
\nonumber
&= \max_{k\in\K}t_{T,k}^d(\RR_d) + \max_{k\in\K}t_{C,k}(\theta,f_k) + \max_{k\in\K}t_{T,k}^u (\theta,\f,\RR_u)
\\
\nonumber
&= t_{d,B}(\RR_d) + \max_{k\in\K}t_{d,k}(R_{d,k})+\max_{k\in\K}t_{C,k}(\theta,f_k)
\\
\nonumber
%+\max_{k\in\K}t_{C,k}(\theta,f_k)
& \qquad + \max_{k\in\K}t_{u,k}(R_{u,k})+ t_{u,B}(\RR_u)
\\
& \triangleq t_{d,B} + t_{d,W} + t_{C} + t_{u,W} + t_{u,B},
\end{align}
\vspace{-5mm}

\noindent
where $\f\triangleq\{f_k\}_{k\in\K}$; $t_{d,W}$ or $t_{u,W}$ is the maximum delay for a complete DLTUT or ULTUT via wireless links; $t_{C}$ is the maximum delay for all the UEs to complete their local training update computation.
Note again that the time of the global training update at the CPU is ignored as discussed in Section~\ref{sec:SystemModel:B}.

As can be seen from \eqref{totaltime:global}, $T_G(\theta,\f,\RR_d,\RR_u)$ relies on both $(\f,\RR_d,\RR_u)$ and $\theta$. However, only $(\f,\RR_d,\RR_u)$ is optimized to reduce the time $T_G(\theta,\f,\RR_d,\RR_u)$ of one iteration of the FL process in each large-scale coherence time.
% in each large-scale coherence time $\widetilde{T}_c$.
This is because any change of $\theta$ leads to the change in the number of iterations $G(\theta)$ of the FL process as shown in \eqref{G}.
Therefore, $(\f,\RR_d,\RR_u)$ and $\theta$ must be optimized independently in different timescales.

To measure how efficient the time of each iteration of the FL process is optimized over different large-scale coherence times, we introduce a new metric termed ``ergodic time of one iteration of the FL process", i.e., $\EEE\{T_G(\theta,\f,\RR_d,\RR_u)\}$.
Here, $\EEE\{T_G(\theta,\f,\RR_d,\RR_u)\}$ is the average of $T_G(\theta,\f,\RR_d,\RR_u)$ over the large-scale fading realizations.
The effective time of one FL process is then defined as
\vspace{-1mm}
\begin{align}\label{totaltime}
\nonumber
T_e(\theta,\f,\RR_d,\RR_u)
&\triangleq G(\theta)\EEE\{T_G(\theta,\f,\RR_d,\RR_u)\}
\\
\nonumber
&= \vartheta\log(\tfrac{1}{\epsilon})
\EEE\Big\{\frac{T_G(\theta,\f,\RR_d,\RR_u)}{1-\theta} \Big\}
\\
&= \vartheta\log(\tfrac{1}{\epsilon})
\EEE\{T(\theta,\f,\RR_d,\RR_u)\},
\end{align}
\vspace{-5mm}

\noindent
where $T(\theta,\f,\RR_d,\RR_u)\triangleq\frac{T_G(\theta,\f,\RR_d,\RR_u)}{1-\theta}$.
For ease of presentation, we make the following definition.
\begin{definition}
An effective training time of FL is the effective time of one FL process and is computed as \eqref{totaltime}.
\end{definition}
The problem of FL training time minimization for the FL framework \cite{ma15OMS} in the considered CFmMIMO system model is thus formulated as:
\vspace{-1mm}
\begin{subequations}\label{mainP}
\begin{align}
\label{CF:mainP}
\underset{\ETA,\ZETA,\theta,\f,\RR_d,\RR_u}{\min} \,\,
&g(\theta,\f,\RR_d,\RR_u)\triangleq\EEE\{T(\theta,\f,\RR_d,\RR_u)\}
\\
\nonumber
\mathrm{s.t.}\,\,
&
%\eqref{cons:d:cap},
\eqref{power:d:cons},\eqref{rate:d}, \eqref{power:u:cons},
%\eqref{cons:u:cap},
\eqref{rate:u}
\\
\label{cons:energy}
& E_{T,k}(\zeta_k,R_{u,k}) + E_{C,k}(\theta,f_k) \leq E_{k,\max}, \forall k
\\
\label{cons:f}
& f_{k,\min}\leq f_{k}\leq f_{k,\max}, \forall k
\\
\label{cons:eta}
& 0\leq \eta_{mk}, \forall m,k
\\
\label{cons:zeta}
& 0\leq \zeta_{k}, \forall k
\\
\label{cons:Rd}
& 0\leq R_{d,k}, \forall k
\\
\label{cons:Ru}
& 0\leq R_{u,k},\forall k
\\
\label{cons:theta}
& \theta_{\min}\leq \theta \leq \theta_{\max}.
\end{align}
\end{subequations}
\vspace{-6mm}

\noindent
Here, problem \eqref{mainP} takes into account the issues related to device performance and user experience, i.e., limiting a maximum energy in \eqref{cons:energy} and a maximum frequency processing \eqref{cons:f} in order to ensure that performing FL does not affect much on the UEs' other functions such as data transmission and computation.
Problem \eqref{mainP} has a nonconvex stochastic, mixed-timescale structure, along with the tight coupling among the variables. Finding its globally optimal solution is challenging. This paper instead aims to propose a solution approach that is suitable for practical implementation.
\begin{remark}
At first glance, the optimization problem \eqref{mainP} is only valid for the FL framework \cite{ma15OMS} because \eqref{L} and \eqref{G}. Nevertheless, on closer observation, the total training time of any FL process (including \cite{ma15OMS}) normally involves variables that are optimized in different timescales. The variables such as local accuracy $\theta$ are optimized in long-term timescales while the variables such as power, rates are optimized in short-term timescales. This leads to the stochastic structure of the training time minimization problems as discussed after \eqref{totaltime:global}.
In this sense, the optimization problems for other FL frameworks can be different from (35) but their stochastic structures are the same as that of \eqref{mainP}. Therefore, we only use the example \cite{ma15OMS} to show this structure. Moreover, the structure of the algorithms to solve these optimization problems is also the same as that shown in the next section.
\end{remark}

\section{FL Training Time Minimization: Proposed Algorithm}
\label{sec:alg}
To resolve problem \eqref{mainP}, we utilize the online successive convex approximation approach for solving two-stage stochastic nonconvex optimization problems in \cite{liu18TSP}. Note that while \cite{liu18TSP} only provides a general description of the solution method, we specifically tailor it to devise a new algorithm for \eqref{mainP}.

According to \cite{Boyd04}, problem \eqref{mainP} can be decomposed into a family of short-term subproblems and a long-term master problem as follows. For a given $\theta$ and large-scale fading coefficients $\BETA\triangleq\{\beta_{mk}\}_{m\in\MM,k\in\K}$ in each large-scale coherence time, the short-term subproblem is expressed as:
\vspace{-2mm}
\begin{align}\label{shortP}
\underset{\ETA,\ZETA,\f,\RR_d,\RR_u}{\min} \,\,
&T(\f,\RR_d,\RR_u)
\\
\nonumber
\mathrm{s.t.}\,\,
&
%\eqref{cons:d:cap},
\eqref{power:d:cons},\eqref{rate:d}, \eqref{power:u:cons},
%\eqref{cons:u:cap},
\eqref{rate:u}, \eqref{cons:energy}-\eqref{cons:Ru}.
\end{align}
\vspace{-6mm}

\noindent
For given optimal solutions $\{(\ETA^*,\ZETA^*,\f^*,\RR_d^*,\RR_u^*)\}$ to problems \eqref{shortP} at all large-scale coherence times, the long-term master problem is expressed as:
\vspace{-1mm}
\begin{align}\label{longP}
%\label{CF:longP}
\underset{\theta}{\min} \,\,
&g(\theta)\triangleq\EEE\{T(\theta)\}
\\
\nonumber
\mathrm{s.t.}\,\,
&
%\eqref{cons:energy},
\eqref{cons:theta}.
\end{align}
\vspace{-12mm}

\noindent
\subsection{Solving the Short-term Subproblem \eqref{shortP}}
\vspace{-0mm}
Problem \eqref{shortP} can be rewritten in an epigraph form as follows.
\begin{subequations}\label{shortP:epi}
\begin{align}
\label{CF:shortP:epi}
\underset{\x}{\min} \,\,
&\frac{\omega}{1-\theta}
\\
\mathrm{s.t.}\,\,
%&  \gamma \geq \max_{k\in\K}t_{T,k}^d(\RR_d) + \max_{k\in\K}t_{T,k}^u (\f,\RR_u)
%\\
\nonumber
& \omega \geq t_{d,B}(\RR_d) + \max_{k\in\K}t_{d,k}(R_{d,k})+\max_{k\in\K}t_{C,k}(\theta,f_k)
\\
&\qquad+ \max_{k\in\K}t_{u,k}(R_{u,k})+ t_{u,B}(\RR_u)
\\
\label{cons:cvx:5}
& \rho_u N_0 \varrho_kS_u + \nu\log\left(\frac{1}{\theta}\right)\frac{\alpha}{2}c_kD_kf_k^2  \leq E_{k,\max}, \forall k
\\
\label{}
& \zeta_k\leq \varrho_kR_{u,k}, \forall k
\\
\nonumber
&
%\eqref{cons:d:cap},
\eqref{power:d:cons},\eqref{rate:d}, \eqref{power:u:cons},
%\eqref{cons:u:cap},
\eqref{rate:u}, \eqref{cons:f}-\eqref{cons:Ru},
\end{align}
\end{subequations}
\vspace{-5mm}

\noindent
where $\x\triangleq(\ETA,\ZETA,\f,\RR_d,\RR_u,\omega,\VARRHO)$, $\omega$ and $\VARRHO\triangleq\{\varrho_k\}_{k\in\K}$ are additional variables. If we let
$\vv \triangleq \{v_{mk}\}_{m\in\MM,k\in\K}$ and $\uu\triangleq \{u_k\}_{k\in\K}$ with
\vspace{-1mm}
\begin{align}\label{v}
v_{mk}&\triangleq \eta_{mk}^{1/2}, \forall m,k,
\\
\label{u}
u_k &\triangleq \zeta_k^{1/2},\forall k,
\end{align}
then \eqref{shortP:epi} can be rewritten as:
\begin{subequations}\label{shortP:equi}
\begin{align}
\label{CF:shortP:equi}
\underset{\widetilde{\x}}{\min} \,\,
&\frac{\omega}{1-\theta}
\\
\mathrm{s.t.}\,\,
\label{cons:cvx:1}
& \omega \geq \frac{KS_d}{\sum_{k\in\K}R_{d,k}} + t_d + t_C + t_u
%\\
%& \qquad
+\frac{KS_u}{\sum_{k\in\K}R_{u,k}}
\\
\label{cons:cvx:2}
& t_d \geq \frac{S_d}{R_{d,k}}, \forall k
\\
\label{cons:cvx:3}
& t_C \geq \frac{\nu\log\left(\frac{1}{\theta}\right)D_kc_k}{f_k}, \forall k
\\
\label{cons:cvx:4}
& t_u \geq \frac{S_u}{R_{u,k}}, \forall k
%\\
%\label{cons:cvx:5}
%& \varrho_kS_u + \nu\log\left(\frac{1}{\theta}\right)\frac{\alpha}{2}c_kD_kf_k^2  \leq E_{k,\max}, \forall k
\\
\label{cons:noncvx:1}
& u_k^2\leq \varrho_kR_{u,k}, \forall k
\\
\label{cons:noncvx:2}
& R_{d,k} \leq h_{d,k}(\vv), \forall k
\\
\label{cons:noncvx:3}
& R_{u,k} \leq h_{u,k}(\uu), \forall k
\\
\label{cons:cvx:6}
& \sum_{k\in\K}\sigma_{mk}^2v_{mk}^2 \leq 1, \forall m
\\
\label{cons:cvx:7}
& 0<v_{mk},\forall m\in\MM,k\in\K
\\
\label{cons:cvx:8}
& 0< u_k\leq 1,\forall k\in\K
\\
\nonumber
&
%\eqref{cons:d:cap},\eqref{cons:u:cap},
\eqref{cons:f}, \eqref{cons:Rd}, \eqref{cons:Ru}, \eqref{cons:cvx:5},
\end{align}
\end{subequations}
\vspace{-7mm}

\noindent
where $\widetilde{\x}\triangleq\{\x,\vv,\uu,t_d,t_C,t_u\}\setminus\{\ETA,\ZETA\}$, $t_d,t_C$ and $t_u$ are additional variables.
Note that
\eqref{shortP:equi} is still challenging due to the nonconvex constraints \eqref{cons:noncvx:1}, \eqref{cons:noncvx:2},  and \eqref{cons:noncvx:3}.

To solve \eqref{shortP:equi}, we first rewrite \eqref{cons:noncvx:1} as
\vspace{-2mm}
\begin{align}
\label{cons:noncvx:3new}
z_k(u_k,\varrho_k,R_{u,k}) \leq 0, \forall k.
\end{align}
\vspace{-6mm}

\noindent
where $z_k(u_k,\varrho_k,R_{u,k})\triangleq 4u_k^2 - (\varrho_k+R_{u,k})^2 + (\varrho_k-R_{u,k})^2$.
%Here, it can be shown that a function $f(x,y)\triangleq(x+y)^2$ is jointly convex in $(x,y)$. Upon applying the first-order Taylor series expansion at a given point $(x^{(n)},y^{(n)})$, the affine lower bound of $f(x,y)$ is given as $2(x^{(n)}+ y^{(n)})(x+y)- (x^{(n)}+ y^{(n)})^2\leq(x+y)^2$ \cite{vu18TWC}.
Note that, for a given point $(x^{(n)},y^{(n)})$, a function $f(x,y)=-(x+y)^2$ has an upper bound $\widetilde{f}(x,y)\geq f(x,y)$ as
\vspace{-2mm}
\begin{align}\label{apprx}
\nonumber
\widetilde{f}(x,y)\triangleq
&-2(x^{(n)}+ y^{(n)})(x+y)+(x^{(n)}+ y^{(n)})^2
\\
&+\delta((x-x^{(n)})^2+(y-y^{(n)})^2),
\end{align}
\vspace{-6mm}

\noindent
where $\delta>0$ can be any constant.
Different from the upper bound used in \cite{vu18TWC}, $\widetilde{f}(x,y)$ is introduced here with the term of $\delta((x-x^{(n)})^2+(y-y^{(n)})^2)$ to ensure its strong convexity.
Now, \eqref{cons:noncvx:1} can be approximated at iteration $n+1$ by the following convex constraint
\vspace{-2mm}
\begin{align}
\label{cons:noncvx:1:appr}
\widetilde{z}_k(u_k,\varrho_k,R_{u,k}) \leq 0, \forall k.
\end{align}
\vspace{-6mm}

\noindent
where
\vspace{-2mm}
\begin{align}\label{z:apprx}
\nonumber
\widetilde{z}_k(u_k,\varrho_k,R_{u,k})
\triangleq
&\,\, 4u_k^2 - 2(\varrho_k^{(n)}
+R_{u,k}^{(n)})(\varrho_k+R_{u,k})
\\
\nonumber
&+ (\varrho_k^{(n)}+R_{u,k}^{(n)})^2
+ (\varrho_k-R_{u,k})^2
\\
&
+\delta((\varrho_k-\varrho_k^{(n)})^2
+(R_{u,k}-R_{u,k}^{(n)})^2).
\end{align}
\vspace{-5mm}

\noindent

To deal with \eqref{cons:noncvx:2} and \eqref{cons:noncvx:3}, we note that a function $f(x,y)= \log\left(1+\frac{|x|^2}{y}\right)$ has the following lower bound \cite{nguyen17TCOM}:
\vspace{-4mm}
\begin{align}\label{}
\nonumber
f(x,y)\geq \log
&\Big(1+\frac{|x^{(n)}|^2}{y^{(n)}}\Big) - \frac{|x^{(n)}|^2}{y^{(n)}} +
\\
&2\frac{x^{(n)}x}{y^{(n)}}
- \frac{|x^{(n)}|^2(|x|^2+y)}{y^{(n)}(|x^{(n)}|^2+y^{(n)})},
%-\delta\big(|x-x^{(n)}|^2+(y-y^{(n)})^2\big),
\end{align}
\vspace{-4mm}

\noindent
where $x\in\RRR, y>0,y^{(n)}>0$.
Therefore, the concave lower bound $\widetilde{h}_{d,k}(\vv)$ of $h_{d,k}(\vv)$ in \eqref{cons:noncvx:2} is given by
\vspace{-2mm}
\begin{align}\label{hd:apprx}
\nonumber
\widetilde{h}_{d,k}(\vv) \triangleq
&\log_2\Big(1+\frac{(\Upsilon_k^{(n)})^2}{\Pi_k^{(n)}}\Big)
-\frac{(\Upsilon_k^{(n)})^2}{\Pi_k^{(n)}}
\\
&+2\frac{\Upsilon_k^{(n)}\Upsilon_k}{\Pi_k^{(n)}}
- \frac{(\Upsilon_k^{(n)})^2(\Upsilon_k^2+\Pi_k)}{\Pi_k^{(n)}((\Upsilon_k^{(n)})^2+\Pi_k^{(n)})}
\leq h_{d,k}(\vv),
\end{align}
\vspace{-3mm}

\noindent
where
\vspace{-2mm}
\begin{align}\label{}
&\Upsilon_k(\{v_{mk}\}_{m\in\MM}) = \sqrt{\rho_d}
\sum_{m\in\MM}v_{mk}\sigma_{mk}^2,
\\
\nonumber
&\Pi_k(\vv) =
\rho_d\sum_{\ell\in\K\setminus k}
\Big(\sum_{m\in\MM}v_{m\ell}\sigma_{m\ell}^2\frac{\beta_{mk}}{\beta_{m\ell}}\Big)^2
|\VARPHI_\ell^H\VARPHI_k|^2
\\
&\qquad\qquad+\rho_d\sum_{\ell\in\K}\sum_{m\in\MM}v_{m\ell}^2\sigma_{m\ell}^2\beta_{mk}+1.
\end{align}
\vspace{-4mm}

\noindent
Similarly, the concave lower bound $\widetilde{h}_{u,k}(\uu)$ of $h_{u,k}(\uu)$ in \eqref{cons:noncvx:3} is given by
\vspace{-2mm}
\begin{align}\label{hu:apprx}
\nonumber
\widetilde{h}_{u,k}(\uu) \triangleq
&\log_2\Big(1+\frac{(\Psi_k^{(n)})^2}{\Xi_k^{(n)}}\Big)
-\frac{(\Psi_k^{(n)})^2}{\Xi_k^{(n)}}+
\\
& 2\frac{\Psi_k^{(n)}\Psi_k}{\Xi_k^{(n)}}
- \frac{(\Psi_k^{(n)})^2(\Psi_k^2+\Xi_k)}{\Xi_k^{(n)}((\Psi_k^{(n)})^2+\Xi_k^{(n)})} \leq h_{u,k}(\uu),
\end{align}
\vspace{-3mm}

\noindent
where
\vspace{-2mm}
\begin{align}\label{}
\Psi_k(u_k) =
&\rho_u^{1/2}u_k(\sum_{m\in\MM}\sigma_{mk}^2),
\\
\nonumber
\Xi_k(\uu) =
&\rho_u\sum_{\ell\in\K\setminus k}u_{\ell}^2
\Big(\sum_{m\in\MM}\sigma_{mk}^2\frac{\beta_{m\ell}}{\beta_{mk}}\Big)^2
|\VARPHI_k^H\VARPHI_\ell|^2
\\
&+\rho_u\sum_{\ell\in\K}u_{\ell}^2\sum_{m\in\MM}\sigma_{mk}^2\beta_{m\ell}
+\sum_{m\in\MM}\sigma_{mk}^2.
\end{align}
\vspace{-4mm}

\noindent
As such, \eqref{cons:noncvx:2} and \eqref{cons:noncvx:3} can be approximated by
\vspace{-2mm}
\begin{align}
\label{cons:noncvx:2:appr}
R_{d,k}\leq \widetilde{h}_{d,k}(\vv), \forall k\in\K,
\\
\label{cons:noncvx:3:appr}
R_{u,k}\leq \widetilde{h}_{u,k}(\uu), \forall k\in\K.
\end{align}
\vspace{-5mm}

\noindent

At the iteration $n+1$, for a given point $\widetilde{\x}^{(n)}$, problem \eqref{shortP:equi} (hence \eqref{shortP}) can finally be approximated by the following convex problem:
\vspace{-2mm}
\begin{align}\label{mainP:appr}
\underset{\widetilde{\x}\in\widetilde{\FF}}{\min} \,\,
\frac{\omega}{1-\theta},
\end{align}
\vspace{-4mm}

\noindent
%\end{subequations}
where $\widetilde{\FF}\triangleq\{
%\eqref{cons:d:cap},
%\eqref{cons:u:cap},
\eqref{cons:f}, \eqref{cons:Rd}, \eqref{cons:Ru}, \eqref{cons:cvx:5},\eqref{cons:cvx:1}-\eqref{cons:cvx:4},
\eqref{cons:cvx:6}-\eqref{cons:cvx:8},
\eqref{cons:noncvx:1:appr},
\eqref{cons:noncvx:2:appr},
\eqref{cons:noncvx:3:appr}\}$ is a convex feasible set.

In Algorithm~\ref{alg:2}, we outline the main steps to solve problem \eqref{shortP}.
Let $\FF\triangleq\{
\eqref{cons:f}, \eqref{cons:Rd}, \eqref{cons:Ru}, \eqref{cons:cvx:5},\eqref{cons:cvx:1}-\eqref{cons:cvx:8}\}$ be the feasible set of \eqref{shortP:equi}.
Starting from a random point $\widetilde{\x}\in\FF$, we solve \eqref{mainP:appr} to obtain its optimal solution $\widetilde{\x}^*$. This solution is then used as an initial point in the next iteration. The algorithm terminates when an accuracy level of $\varepsilon$ is reached.
%The converged solution of Algorithm~\ref{alg:2} will fulfill the KKT conditions of the main problem \eqref{shortP}.
It can be confirmed that $\widetilde{h}_{d,k}(\vv)$ and $\widetilde{h}_{u,k}(\uu)$ satisfy the key properties of general inner approximation functions \cite[Properties (i), (ii), and (iii)]{Marks78OR}.
In the case when the feasible set of problem \eqref{mainP:appr} satisfies Slater's constraint qualification condition, Algorithm~\ref{alg:2} converges to a Karush-Kuhn-Tucker (KKT) solution of \eqref{shortP:equi} when starting from a point $\widetilde{\x}^{(0)}\in\FF$ \cite[Theorem 1]{Marks78OR}. In the worse case when the feasible set of problem \eqref{mainP:appr} does not satisfy Slater's constraint qualification condition, Algorithm~\ref{alg:2} converges to a Fritz John (FJ) solution of \eqref{shortP:equi} \cite[Proposition 2]{vu18TWC}.
By using the variable transformations \eqref{v} and \eqref{u}, it can be seen that the KKT (respectively, FJ) solutions of \eqref{shortP:equi} satisfy the KKT (respectively, FJ) conditions of \eqref{shortP:epi} as well as of \eqref{shortP}.
\begin{algorithm}[!t]
\caption{Solving the short-term subproblem \eqref{shortP}}
\begin{algorithmic}[1]\label{alg:2}
%\small{
\STATE \textbf{Initialization}: Set $n=1$ and choose a random point $\widetilde{\x}^{(0)}\in\FF$.
\REPEAT
\STATE Update $n=n+1$
\STATE Solving \eqref{mainP:appr} to get its optimal solution $\widetilde{\x}^*$
\STATE Update $\widetilde{\x}^{(n)}=\widetilde{\x}^*$
\UNTIL{convergence}
%}
\end{algorithmic}
\textbf{Output}: $(\ETA^*,\ZETA^*,\f^*,\RR_d^*,\RR_u^*)$
\end{algorithm}

\subsection{Solving the Long-term Master Problem \eqref{longP}}
At the large-scale coherence time $n+1$, we replace the cost function of the stochastic nonconvex problem \eqref{longP} by a sample surrogate function as \cite{liu18TSP}
\vspace{-1mm}
\begin{align}\label{sfunc}
\widetilde{g}^{(n+1)}(\theta) = (1-\phi^{(n+1)})\widetilde{g}^{(n)}(\theta) + \phi^{(n+1)}\widetilde{T}(\theta),
\end{align}
\vspace{-6mm}

\noindent
where $\phi^{(n+1)}$ is a weighting parameter. $\widetilde{g}^{(n+1)}(\theta)$ depends on the surrogate function $\widetilde{g}^{(n)}(\theta)$ of the previous large-scale coherence time $(n)$ and the approximate function $\widetilde{T}(\theta)$ of $T(\theta)$.
Here, $\widetilde{g}^{(n)}(\theta)$ is  approximately updated as
\vspace{-1mm}
\begin{align}\label{tildeg}
\widetilde{g}^{(n)}(\theta)
= g^{(n)} +(\nabla g)^{(n)}(\theta-\theta^{(n+1)}),
%+ \tau(\theta-\theta^{(n)})^2
\end{align}
\vspace{-6mm}

\noindent
and $\widetilde{T}(\theta)$ is expressed as
\vspace{-2mm}
\begin{align}\label{tildeT}
\widetilde{T}(\theta) = T^{(n+1)}+(\nabla T)^{(n+1)}(\theta-\theta^{(n+1)})+\tau(\theta-\theta^{(n+1)})^2,
\end{align}
\vspace{-4mm}

\noindent
where $\tau>0$ can be any constant.

With \eqref{tildeg} and \eqref{tildeT},  \eqref{sfunc} can be rewritten as:
\vspace{-1mm}
\begin{align}\label{sfunc:rew}
\nonumber
\widetilde{g}^{(n+1)}(\theta) =
&\,\, g^{(n+1)} + (\nabla g)^{(n+1)}(\theta-\theta^{(n+1)})
\\
&+ \tau(\theta-\theta^{(n+1)})^2,
\end{align}
\vspace{-6mm}

\noindent
where $g^{(n+1)}$ and $(\nabla g)^{(n+1)} $ are updated as
\vspace{-1mm}
\begin{align}\label{}
g^{(n+1)}
&= (1-\phi^{(n+1)})(g)^{(n)} + \phi^{(n+1)} T^{(n+1)}
\\
(\nabla g)^{(n+1)}
&= (1-\phi^{(n+1)})(\nabla g)^{(n)} + \phi^{(n+1)} (\nabla T)^{(n+1)},
\end{align}
\vspace{-4mm}

\noindent
with $g^{(0)}=0$ and $(\nabla g)^{(0)} = 0$. Here,
\vspace{-1mm}
\begin{align}\label{}
(\nabla T)^{(n+1)}=\frac{a+b\log(1/\theta^{(n+1)})-b(1/\theta^{(n+1)}-1)}{(1-\theta^{(n+1)})^2},
\end{align}
\vspace{-4mm}

\noindent
where $a = t_{d,B}^{(n+1)} + t_{d,W}^{(n+1)} + t_{u,W}^{(n+1)} + t_{u,B}^{(n+1)}$ and $b=\nu \, \underset{k}{\max}\left(\frac{D_kc_k}{f_k}\right)$.
Since $\widetilde{g}^{(n+1)}(\theta)$ in \eqref{sfunc:rew} approximates $g(\theta)$ in \eqref{longP}, problem \eqref{longP} is finally approximated by the following convex problem:
\vspace{-2mm}
\begin{align}\label{longP:appr}
\underset{\theta}{\min} \,\,
&\{g^{(n+1)} + (\nabla g)^{(n+1)}(\theta-\theta^{(n+1)})
+ \tau(\theta-\theta^{(n+1)})^2\}
\\
\nonumber
\mathrm{s.t.}\,\,
&
\eqref{cons:theta}.
\end{align}
\vspace{-11mm}

\noindent
\subsection{Solving the Overall Problem \eqref{mainP}}
\label{subsec:alg3}
\vspace{-1mm}

Algorithm~\ref{alg:main} outlines the main steps to solve the overall problem \eqref{mainP}. In the large-scale coherence time $n$, a random large-scale fading coefficient $\BETA$ is realized. For a given random value of $\theta^{(n+1)}\in(0,1)$, one short-term subproblem \eqref{shortP} is solved by Algorithm~\ref{alg:2} after $I^{(n)}$ iterations to obtain a KKT solution.
This solution is then used to construct the approximate long-term master problem \eqref{longP:appr}. After solving \eqref{longP:appr} to obtain an optimal solution $\theta^*$, we update $\theta^{(n+2)}$ as
\vspace{-1mm}
\begin{align}\label{theta}
\theta^{(n+2)} = (1-\pi^{(n+1)})\theta^{(n+1)}+\pi^{(n+1)}\theta^*,
\end{align}
\vspace{-6mm}

\noindent
where $\pi^{(n+1)}$ is a weighting parameter.
Here, $\{\phi^{(n)},\pi^{(n)}\}$ is chosen to satisfy the following conditions \cite[Assumption 5]{liu18TSP}.
\vspace{-5mm}

\noindent
\begin{description}
  \item[(C1):] $\phi^{(n)}\rightarrow 0$, $\frac{1}{\phi^{(n)}}\leq \OO (n^\varsigma)$ for $\varsigma\in(0,1)$, and $\sum_{n}(\phi^{(n)})^2<\infty$;
  \item[(C2):] $\pi^{(n)}\rightarrow 0$, $\sum_{n}\pi^{(n)}=\infty$, $\sum_{n}(\pi^{(n)})^2<\infty$, and $\lim_{n\rightarrow\infty}\frac{\pi^{(n)}}{\phi^{(n)}}=0$.
\end{description}
\vspace{-5mm}

\noindent
\subsection{The Proposed Algorithm: Implementation and Convergence Analysis}
\vspace{-1mm}

\noindent
Referring to Fig.~\ref{fig:time1}, Algorithm~\ref{alg:main} is executed in the ``FL performance optimization'' interval. Specifically, Steps $3$ and $4$ takes place in the time block of STO, while steps $5-8$ in the time block of LTO.
%One iteration of Algorithm~\ref{alg:main} is within one large-scale coherence time $\widetilde{T}_c$.
Once Algorithm~\ref{alg:main} converges, the FL process is then executed using the value of $\theta$ given by Algorithm~\ref{alg:main}.
Here, the performance of training update transmission in each iteration of the FL process is enhanced by updating $(\ETA,\ZETA,\f,\RR_d,\RR_u)$ using  Algorithm~\ref{alg:2} in the STO time block.
Whenever the statistics of large-scale fading changes, Algorithm~\ref{alg:main} is executed again to make sure the FL performance is optimized with the updated statistics.

The convergence of Algorithm~\ref{alg:main} is proved as follows. From the definitions of $\widetilde{z}_k(\varrho_k,R_{u,k})$, $\widetilde{h}_{d,k}(\vv)$, and $\widetilde{h}_{u,k}(\uu)$ in \eqref{z:apprx}, \eqref{hd:apprx} and \eqref{hu:apprx}, it can be verified that $\widetilde{z}_k(\varrho_k,R_{u,k})$, $\widetilde{h}_{d,k}(\vv)$ and $\widetilde{h}_{u,k}(\uu)$ have the following properties:
\begin{itemize}
  \item  $\widetilde{z}_k(\varrho_k^{(n)},R_{u,k}^{(n)})={z}_k(\varrho_k^{(n)},R_{u,k}^{(n)})$,
  $\widetilde{h}_{d,k}(\vv^{(n)})=h_{d,k}(\vv^{(n)})$, $\widetilde{h}_{u,k}(\uu^{(n)})={h}_{u,k}(\uu^{(n)})$,
  $\nabla\widetilde{z}_k(\varrho_k^{(n)},R_{u,k}^{(n)})=\nabla{z}_k(\varrho_k^{(n)},R_{u,k}^{(n)})$,
      $\nabla\widetilde{h}_{d,k}(\vv^{(n)})=\nabla h_{d,k}(\vv^{(n)})$, $\nabla\widetilde{h}_{u,k}(\uu^{(n)})=\nabla{h}_{u,k}(\uu^{(n)})$;
  \item  $\widetilde{z}_k(\varrho_k,R_{u,k})$, $-\widetilde{h}_{d,k}(\vv)$, and  $-\widetilde{h}_{u,k}(\uu)$ are strongly convex;
  \item $\widetilde{z}_k(\varrho_k,\varrho_k^{(n)},R_{u,k},R_{u,k}^{(n)})$ is Lipschitz continuous in all $\varrho_k,\varrho_k^{(n)},R_{u,k},R_{u,k}^{(n)}$;
  $\widetilde{h}_{d,k}(\vv,\vv^{(n)})$ and $\widetilde{h}_{u,k}(\uu,\uu^{(n)})$ are Lipschitz continuous in both $\vv,\vv^{(n)}$ and both $\uu,\uu^{(n)}$, respectively.
\end{itemize}
\vspace{-1mm}

\noindent
Algorithm~\ref{alg:main} thus satisfies all the conditions for the short-term algorithm to work, as specified in the general framework \cite[Assumption 2]{liu18TSP}. As such, the convergence of Algorithm~\ref{alg:main} to a stationary point of problem \eqref{mainP} is guaranteed if $I^{(n)}\rightarrow \infty$ and $N\rightarrow \infty$ \cite[Theorem 2]{liu18TSP}, where the FJ condition may replace the KKT condition in the definition of the stationary point \cite[Definition 1]{liu18TSP}.
In practice, since there are always numerical errors in computation, it is acceptable to choose finite $\{I^{(n)}\}_{n\in\NN}$ and $N$, where $\NN\triangleq\{1,...,N\}$. Therefore, Algorithm~\ref{alg:main} is then guaranteed to converge to the neighbourhood of the stationary solutions of problem \eqref{mainP} \cite[Theorem 3]{liu18TSP}.

The REPEAT-UNTIL loop runs for $N$ iterations before Algorithm~\ref{alg:main} converges.
\vspace{-4mm}

\noindent
\section{Cell-Free TDMA Massive MIMO and Collocated Massive MIMO for Wireless Federated Learning}
\label{sec:TDMACOL}
For comparison, this section introduces cell-free TDMA massive MIMO and collocated massive MIMO approaches that support wireless FL. Their associated problem formulations and solution algorithms are discussed in the following.
\vspace{-5mm}

\noindent
\subsection{Cell-Free TDMA Massive MIMO}
\vspace{-1mm}

\noindent
The channel estimation of cell-free TDMA massive MIMO networks is equivalent to that of the CFmMIMO networks where all the pilot are pairwisely orthogonal, i.e., $\VARPHI_{\ell}^H\VARPHI_{k}=0, \forall \ell\in\K\setminus k$. While cell-free TDMA massive MIMO networks only require the length of the pilot sequence $\widetilde{\tau}_t$ to be $1$, CFmMIMO networks require $\tau_t\geq K$ for orthogonal pilots with $K$ being the number of UEs.

\begin{algorithm}[!t]
\caption{Training time minimization for FL on CFmMIMO networks}
\begin{algorithmic}[1]\label{alg:main}
%\small{
\STATE \textbf{Initialization}: Set $n=0$ and choose a random point $\theta^{(n+1)}\in(0,1)$.
\REPEAT
\STATE A random $\BETA$ is realized for one large-scale coherence time
\STATE Find the optimal solution $(\ETA^*,\ZETA^*,\f^*,\RR_d^*,\RR_u^*)$ of the short-term subproblem \eqref{shortP} by using Algorithm~\ref{alg:2}
\STATE Update $(\ETA^{(n+1)},\ZETA^{(n+1)},\f^{(n+1)},\RR_d^{(n+1)},\RR_u^{(n+1)})=
(\ETA^*,\ZETA^*,\f^*,\RR_d^*,\RR_u^*)$
\STATE Solve the approximate long-term master problem \eqref{longP:appr} to obtain its optimal solution $\theta^*$
\STATE Update $\theta^{(n+2)}$ by \eqref{theta}
\STATE Update $n=n+1$
\UNTIL{convergence}
%}
\end{algorithmic}
\textbf{Output}: $\theta^*=\theta^{(n+1)}$
\end{algorithm}

In cell-free TDMA massive MIMO networks, the training update transmissions between the APs and $K$ UEs
happen in $K$ equal orthogonal time slots. Therefore, a factor of $(1/K)$ is imposed on the achievable DL and UL rates. Specifically, the achievable DL rate for a UE $k$ is
\vspace{-2mm}
\begin{align}\label{rate:d:TDMA}
R_{d,k} \leq
\frac{\tau_c-\widetilde{\tau}_{t}}{K\tau_c}B\log_2\Big(1+\frac
{\rho_p\big(\sum_{m\in\MM}\eta_{mk}^{1/2}\sigma_{mk}^2\big)^2}
{\rho_p\sum_{m\in\MM}\eta_{mk}\sigma_{mk}^2\beta_{mk}+1}\Big),
\end{align}
\vspace{-4mm}

\noindent
where $\sigma_{mk}^2=\frac{\widetilde{\tau}_{t}\widetilde{\rho}_{t}(\beta_{mk})^2}
{\widetilde{\tau}_{t}\widetilde{\rho}_{t}\beta_{mk}+1}$, and $\widetilde{\rho}_t$ is the normalized signal-to-noise ratio (SNR) of each pilot symbol. The achievable UL rate $R_{u,k}$ for a UE $k$ is
\vspace{-2mm}

\noindent
\begin{align}\label{rate:u:TDMA}
R_{u,k} \leq
\frac{\tau_c-\widetilde{\tau}_{t}}{K\tau_c}B\log_2\Big(
1+
\tfrac
{\rho_u\zeta_{k}\left(\sum_{m\in\MM}\sigma_{mk}^2\right)^2}
{\rho_u\zeta_{k}\sum_{m\in\MM}\sigma_{mk}^2\beta_{mk}
+\sum_{m\in\MM}\sigma_{mk}^2}
\Big).
\end{align}
\vspace{-4mm}

\noindent
Since the training updates are transmitted sequentially via wireless links, the effective training time of FL in cell-free TDMA massive MIMO networks is expressed as
\vspace{-2mm}
\begin{align}\label{totaltime:TDMA}
\nonumber
&T_{\text{e,TDMA}}(\theta,\f,\RR_d,\RR_u)
\triangleq
G(\theta)\EEE\{T_{\text{G,TDMA}}(\theta,\f,\RR_d,\RR_u)\}
\\
\nonumber
\triangleq& \vartheta\log\left(\tfrac{1}{\epsilon}\right)
\EEE\Big\{\tfrac{1}{1-\theta}\Big(t_{d,B}(\RR_d) + \sum_{k\in\K}t_{d,k}(R_{d,k})
\\
\nonumber
&\qquad+ \max_{k\in\K}t_{C,k}(\theta,f_k) + \sum_{k\in\K}t_{u,k}(R_{u,k}) + t_{u,B}(\RR_u) \Big)\Big\}
\\
\triangleq
&\vartheta\log\left(\tfrac{1}{\epsilon}\right)
\EEE\{T_{\text{TDMA}}(\theta,\f,\RR_d,\RR_u)\}.
\end{align}
\vspace{-6mm}

\noindent
The problem of FL training time minimization for cell-free TDMA massive MIMO is formulated as:
\vspace{-2mm}
\begin{subequations}\label{mainP:TDMA}
\begin{align}
\label{CF:mainP:TDMA}
\underset{\ETA,\ZETA,\theta,\f,\RR_d,\RR_u}{\min} \,\,
&\EEE\{T_{\text{TDMA}}(\theta,\f,\RR_d,\RR_u)\}
\\
\nonumber
\mathrm{s.t.}\,\,
\nonumber
&
%\eqref{cons:d:cap},
\eqref{power:u:cons},
%\eqref{cons:u:cap},
\eqref{cons:energy}-\eqref{cons:theta}, \eqref{rate:d:TDMA}, \eqref{rate:u:TDMA}
\\
& \sigma_{mk}^2\eta_{mk} \leq 1, \forall m.
\end{align}
\end{subequations}
\vspace{-6mm}

\noindent
Since problem \eqref{mainP:TDMA} has the same mathematical structure as \eqref{mainP}, the former can be solved by a slightly modified version of Algorithm~\ref{alg:main} proposed in Section~\ref{sec:alg}.
\vspace{-4mm}

\noindent
\subsection{Collocated Massive MIMO}
\vspace{-1mm}

A collocated massive MIMO network is a special case of a CFmMIMO network where all the APs are collocated. Therefore, $\beta_{mk}=\beta_k$ and $\sigma_{mk}^2=\sigma_{k}^2,\forall k\in\K$. The DL power control coefficient $\eta_{k},\forall k\in\K$, is constrained by
\vspace{-2mm}

\noindent
\begin{align}\label{power:d:col:cons}
\sum_{k\in\K}\sigma_{k}^2\frac{\eta_{k}}{M} \leq 1.
\end{align}
\vspace{-4mm}

\noindent
From \eqref{rate:d} and \eqref{rate:u}, the achievable DL and UL rates for UE $k$ are respectively designed as \eqref{rate:d:col} and \eqref{rate:u:col} at the top of the next page.
\begin{figure*}
\begin{align}\label{rate:d:col}
R_{d,k}
&\leq\frac{\tau_c-\tau_t}{\tau_c}
B\log_2\Big(1+\frac
{\rho_pM\eta_{k}\sigma_{k}^4}
{\rho_p\sum_{\ell\in\K\setminus k}
M\eta_{\ell}\left(\sigma_{\ell}^2\frac{\beta_{k}}{\beta_{\ell}}\right)^2
|\VARPHI_\ell^H\VARPHI_k|^2
+\rho_p\sum_{\ell\in\K}\eta_{\ell}\sigma_{\ell}^2\beta_{k}+1}
\Big)
%\end{align}
%\begin{align}
\\
\label{rate:u:col}
R_{u,k}
&\leq \frac{\tau_c-\tau_t}{\tau_c}B\log_2\Big(
1+
\frac
{\rho_uM\zeta_{k}\sigma_{k}^2}
{\rho_u\sum_{\ell\in\K\setminus k}\zeta_{\ell}M\sigma_{k}^2
\left(\frac{\beta_{\ell}}{\beta_{k}}\right)^2
|\VARPHI_k^H\VARPHI_\ell|^2
+\rho_u\sum_{\ell\in\K}\zeta_{\ell}\beta_{\ell}
+1}
\Big).
\end{align}
\hrulefill
\end{figure*}

The problem of FL training time minimization for collocated massive MIMO is formulated as:
\vspace{-2mm}
\begin{align}\label{mainP:col}
\underset{\ETA,\ZETA,\theta,\f,\RR_d,\RR_u}{\min} \,\,
&\EEE\{T(\theta,\f,\RR_d,\RR_u)\}
\\
\nonumber
\mathrm{s.t.}\,\,
\nonumber
&
\eqref{power:u:cons},
\eqref{cons:energy}-\eqref{cons:theta},
\eqref{power:d:col:cons}-\eqref{rate:u:col}.
\end{align}
\vspace{-7mm}

\noindent
Similar to \eqref{mainP:TDMA}, problem \eqref{mainP:col} can be solved by a slightly modified version of Algorithm~\ref{alg:main} in Section~\ref{sec:alg}.
\vspace{-3mm}

\noindent
\section{Numerical Examples}
\label{sec:sim}
\subsection{Parameters and Setup}
We consider a CFmMIMO network with $\tau_c=200$ samples. The APs and UEs are located in a square of $D\times D$ km$^2$ whose edges are wrapped around to avoid the boundary effects. The large-scale fading coefficients, e.g., $\beta_{mk}$, are modeled in the same manner as \cite{ngo18TGN}: $\beta_{mk} = 10^{\frac{\text{PL}_{mk}^d}{10}}10^{\frac{\sigma_{shd}z_{mk}^d}{10}}$,
where $10^{\frac{\sigma_{shd}z_{mk}^d}{10}}$ represents the log-normal shadowing with the standard deviation $\sigma_{shd}$ (in dB); and $10^{\frac{\text{PL}_{mk}^d}{10}}$ represents the three-slope path loss. $\text{PL}_{mk}^d$ (in dB) is given by
\vspace{-2mm}
\begin{align}\label{PL:model}
\text{PL}_{mk}^d =
\scriptsize{
\begin{cases}
  -L-35\log_{10}(d_{mk}), &\text{if $d_{mk}>d_1$}, \\
  -L-15\log_{10}(d_1)-20\log_{10}(d_{mk}), &\text{if $d_0<d_{mk}\leq d_1$},\\
  -L-15\log_{10}(d_1)-20\log_{10}(d_0), &\text{if $d_{mk}\leq d_0$},
\end{cases}
}
\end{align}
\vspace{-4mm}

\noindent
where $L$ is a constant depending on the carrier frequency, the UE and AP heights.
To estimate channels, a scheme of random pilot is used in the time block of UL channel estimation. Specifically, the pilot of each user is randomly chosen from a predefined set of $\tau_t$ orthogonal pilot sequences of length $\tau_t$ samples.

Here, we choose $\sigma_{shd}=8$ dB, $d_0=10$ m, $d_1=50$ m, $L=140.7$ dB, bandwidth $B = 20$ MHz, noise figure $F=9$ dB \cite{ngo17TWC}, $f_{k,\max}=f_{\max}=3.0\times 10^9$ cycles/s, $f_{k,\min}=f_{\min}=1\times 10^6$ cycles/s, $D_k=\widehat{D}=10$ MB, $c_k=c=20$ cycles/sample,
$\forall k$, $\nu=\vartheta=1$, $S_d=S_u=5$ MB, $\alpha=2\times 10^{-28}$ \cite{tran19INFOCOM},
$E_{k,\max}=E_{\max}=15$ J, $\theta_{\max} = -10$ dB,
$\theta_{\min} = -60$ dB,
and $\epsilon=\varepsilon=10^{-2}$.
Noise power $N_0=k_BT_0BF=-92$ dBm, where $k_B=1.381\times10^{-23}$ Joules/$^o$K is the Boltzmann constant and $T_0 = 290\,^o$K is the noise temperature.
Let $\tilde{\rho}_d=1$ W, $\tilde{\rho}_u=0.2$ W and $\tilde{\rho}_t=0.2$ W be the maximum transmit power of the APs, UEs and UL pilot sequences, respectively. The maximum transmit powers $\rho_d$, $\rho_u$ and $\rho_t$ are normalized by the noise power. We set $\pi^{(n)} = \frac{1}{n}$ and $\phi^{(n)} = \frac{1}{n^{7/8}}$ which satisfy conditions (C1) and (C2) in Section~\ref{subsec:alg3}.
\vspace{-1mm}

\noindent
\begin{remark}
Our paper does not propose a new FL framework but rather a scheme for a CFmMIMO network to support any FL framework. Here, we consider an existing FL framework \cite{ma15OMS} as an example to show how to improve the FL performance in terms of training time minimization. Therefore, the simulations on real datasets to see the effectiveness of the considered FL framework has already been performed in \cite[Section 6]{ma15OMS}, and hence, they are not considered in this paper. Instead, in what follows, we focus on the numerical results to analyze the effectiveness of the proposed Algorithm~\ref{alg:main} to solve the problem \eqref{mainP} of FL training time minimization for the FL framework \cite{ma15OMS}.
\end{remark}
\begin{figure}[t!]
\centering
\includegraphics[width=0.4\textwidth]{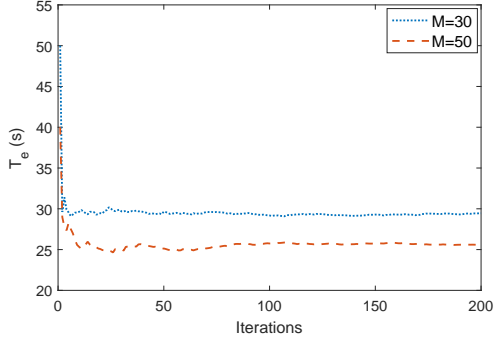}
\caption{The convergence of Algorithm~\ref{alg:main}. Here, $K=4$.}
\label{fig:1}
\end{figure}
\vspace{-6mm}

\noindent
\subsection{Results and Discussions}
\vspace{-1mm}

\noindent
\subsubsection{Effectiveness of the proposed algorithm}
First, we evaluate the convergence behavior of the proposed Algorithm~\ref{alg:main}. Fig.~4 shows the effective training time $T_e$ versus the number of iterations with $D=0.5$ km, $K=4,\tau_t=10$ and $M=\{30,50\}$ for an arbitrary large-scale fading realization. It can be seen from Fig.~4 that Algorithm~\ref{alg:main} converges in fewer than $100$ iterations. It is also worth noting that each iteration of Algorithm~\ref{alg:main} corresponds to solving simple convex programs \eqref{mainP:appr} and \eqref{longP:appr}. It is therefore expected that  Algorithm~\ref{alg:main} has a low computational complexity.

To further evaluate the effectiveness of  Algorithm~\ref{alg:main}, we consider the following baseline schemes:
\begin{itemize}
\item Baseline 1 (BL1): The DL powers allocated to all UEs are the same, i.e., $\eta_{mk}\sigma_{mk}^2=1/K,\forall m,k$. The transmitted UL power of each UE is maximum, i.e., $\zeta_k=1,\forall k$. The local accuracy is fixed, i.e., $\theta = \frac{\theta_{\max}+\theta_{\min}}{2}$ dB. The data rates and processing frequencies of UEs are then optimized.
\item Baseline 2 (BL2): This baseline is similar to BL1 except that $\theta$ is optimized by a slightly modified version of Algorithm~\ref{alg:main} (without using Algorithm~\ref{alg:2}).
\item Baseline 3 (BL3): This baseline is similar to BL1 except that the transmitted DL and UL powers are optimized by Algorithm~\ref{alg:2}. Here, the effective training time of FL is the averaged time of one FL process taken over the large-scale fading realizations.
\end{itemize}

\begin{figure}[t!]
\centering
\begin{minipage}[t]{0.4\textwidth}
\includegraphics[width=1\textwidth]{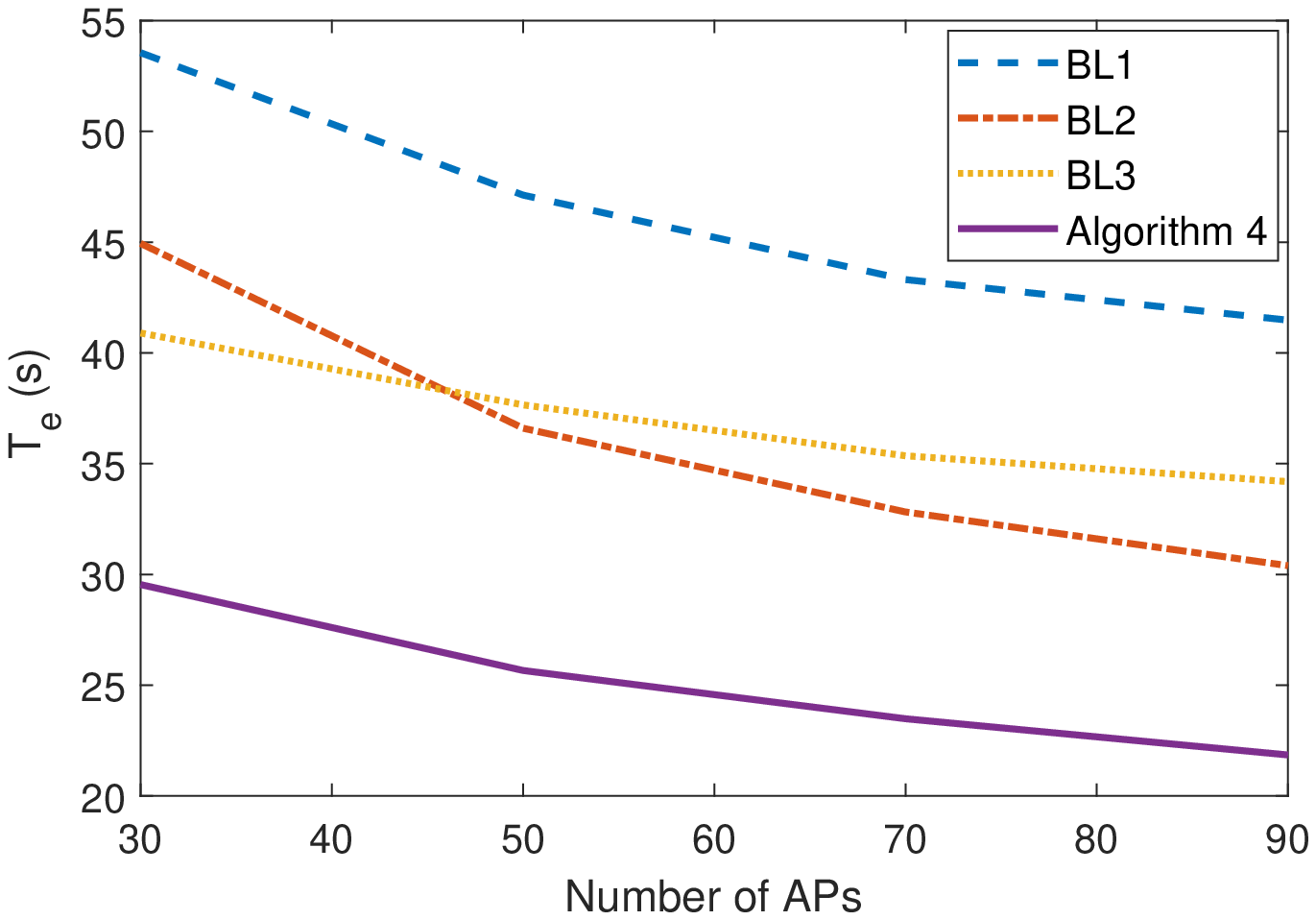}
\caption{Comparison among the baselines and Algorithm~\ref{alg:main}. %with the comparison among the baselines BL1, BL2, BL3 and the proposed algorithm.
Here, $K=4$.}
\label{fig:2}
\end{minipage}
\begin{minipage}[t]{0.4\textwidth}
\includegraphics[width=1\textwidth]{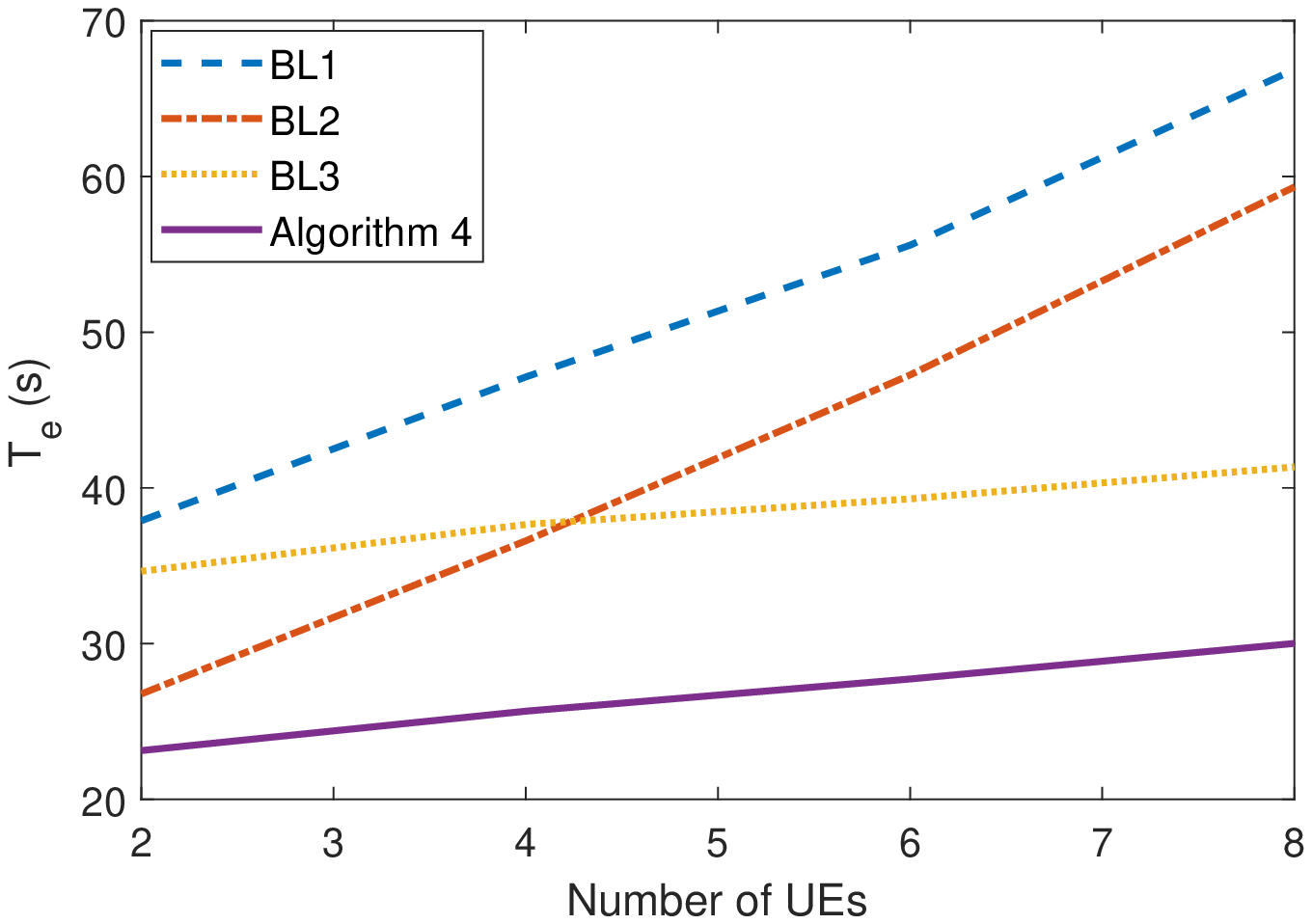}
\caption{Comparison among the baselines and Algorithm~\ref{alg:main}. %with the comparison among the baselines BL1, BL2, BL3 and the proposed algorithm.
Here, $M=50$.}
\label{fig:3}
\end{minipage}
\end{figure}

Figs.~\ref{fig:2} and~\ref{fig:3} compare the  effective training time $T_e$ by the considered schemes.
%Fig.~\ref{fig:2} displays the effective time as a function of the number of APs, while Fig.~\ref{fig:3} displays the effective time as a function of the number of UEs.
As seen, Algorithm~\ref{alg:main} gives the best performance. In particular, compared to BL1, the time reduction by Algorithm~\ref{alg:main} is up to $55\%$ with $M=50$, $K=8$.
Note that BL2 and BL3 also perform much better than BL1, e.g., up to $29\%$ in term of time reduction with $M=50, K=2$ and $38\%$ with $M=50, K=8$, respectively. Even so, Algorithm~\ref{alg:main} still provides substantial time reductions over BL2 and BL3, e.g., up to $49\%$ with $M=50,K=8$ and $43\%$ with $M=90,K=4$, respectively.

The figures not only show the importance of optimizing transmit power or local accuracy, but also demonstrate the noticeable advantage of joint optimization design. Moreover, thanks to the array gain, the data rates of UEs increase when the number of APs increases. This leads to the decrease in the effective training time as shown in Fig.~\ref{fig:2}.
It can also be observed from Fig.~\ref{fig:3} that a dramatic increase in the training time when the number of UEs increases. This is because the mutual interference and pilot contamination become stronger for a larger number of UEs.
%\begin{remark}
%The figures also show that the training time is in the order of dozens of seconds. Intuitively, this is reasonable because as shown in \eqref{totaltime}, the training time of one FL process is the product of the number $G$ of global iterations and the time $T_G$ of one iteration, where $G$ is in the order of $10$ and $T_G$ is in the order of seconds.
%\end{remark}

%\subsubsection{AP selection}
%y-axis: training time, x-axis: the number of APs. Different number of backhaul capacities.

\begin{figure}[t!]
\centering
\begin{minipage}[t]{0.4\textwidth}
\includegraphics[width=1\textwidth]{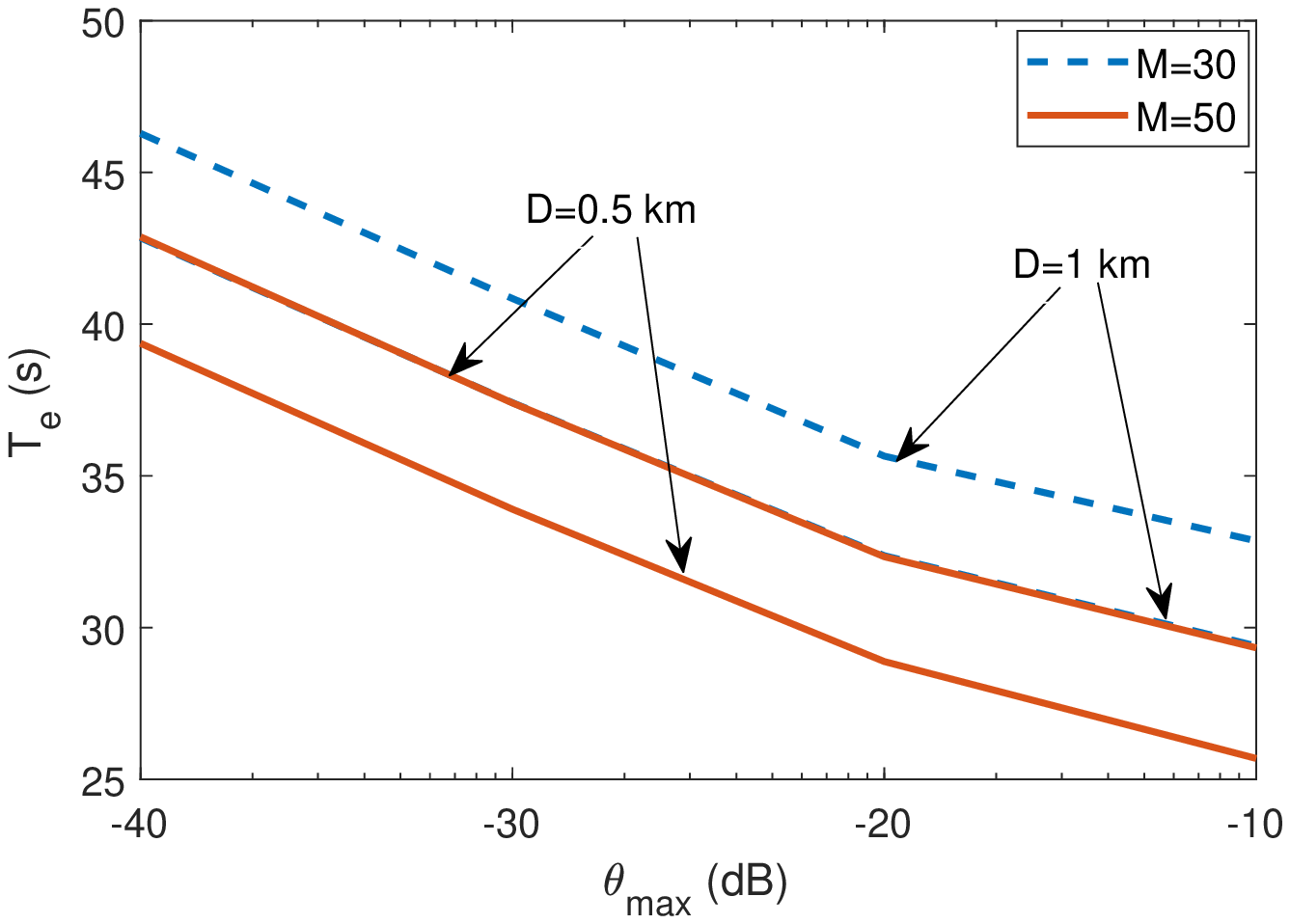}
\caption{Impact of the local accuracy on the effective training time. Here, $K=4$.}
\label{fig:4}
\end{minipage}
\begin{minipage}[t]{0.4\textwidth}
\includegraphics[width=1\textwidth]{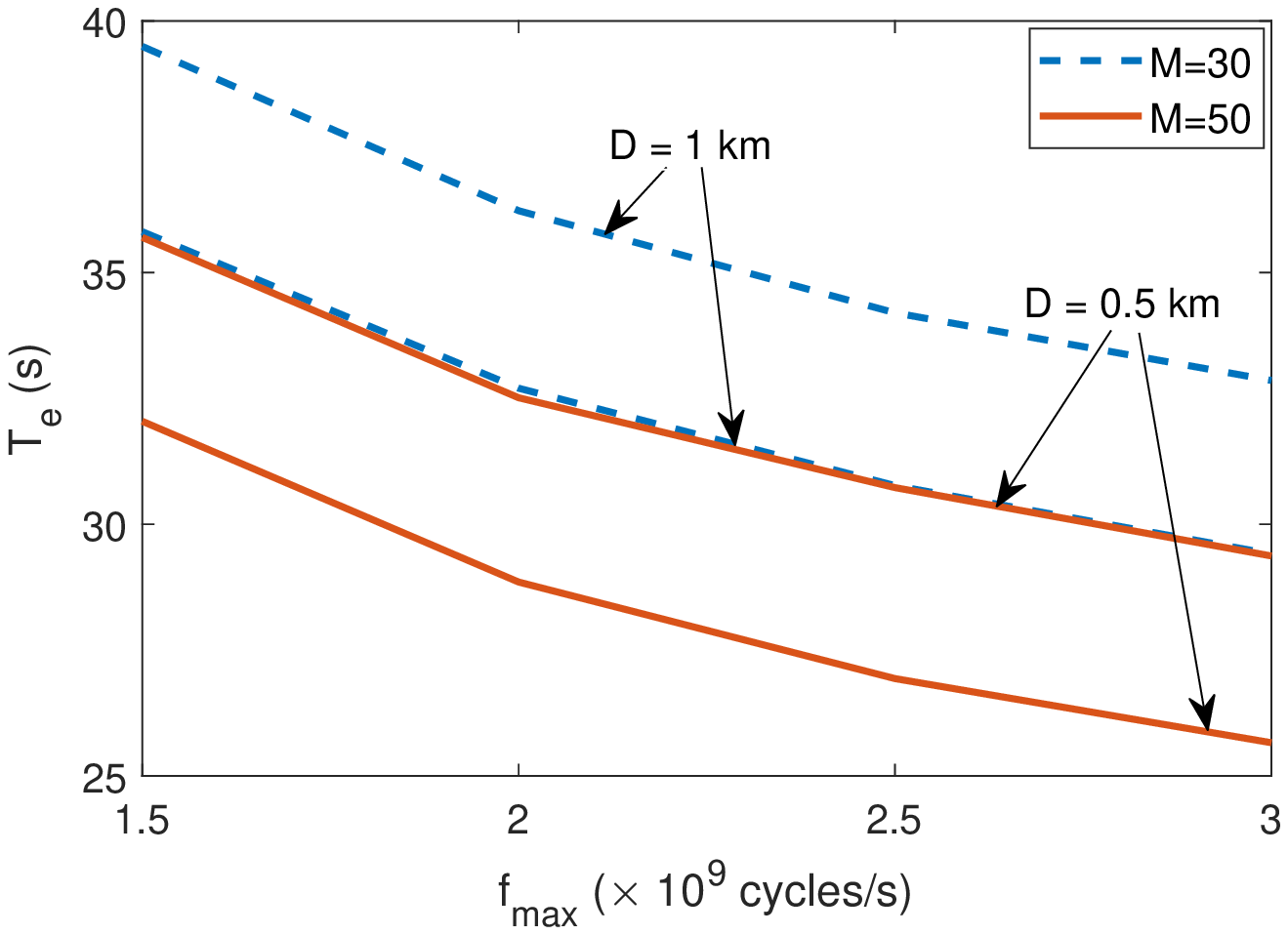}
\caption{Impact of UE's processing frequency on the effective training time. Here, $K=4$.}
\label{fig:5}
\end{minipage}
\end{figure}

\subsubsection{Impact of key system parameters on the effective training time}
The impact of the local accuracy on the effective training time is shown in Fig.~\ref{fig:4}. Decreasing the threshold $\theta_{\max}$ leads to a dramatic increase in the effective training time, e.g., by up to $33\%$ with $\theta_{\max}=-40$ dB in comparison to that with $\theta_{\max}=-10$. This is reasonable because at a lower value of $\theta$, more iterations are required for local training. To keep the energy consumption of UEs below $E_{\max}$, the UEs' processing frequencies become smaller. This leads to an increase in the time required to compute the local training updates.

Fig.~\ref{fig:5} shows the impact of UE's processing frequency on the effective training time. As seen, the effective training time increases when the threshold $f_{\max}$ decreases. In particular, the increase is by up to $19\%$ with $f_{\max}=1.5\times 10^9$ cycles/s in comparison to that with $f_{\max}=3\times 10^9$ cycles/s.
This is because at a lower value of UEs' processing frequency, it requires more time to compute the local training updates.

\begin{figure}[t!]
\centering
\begin{minipage}[t]{0.4\textwidth}
\includegraphics[width=1\textwidth]{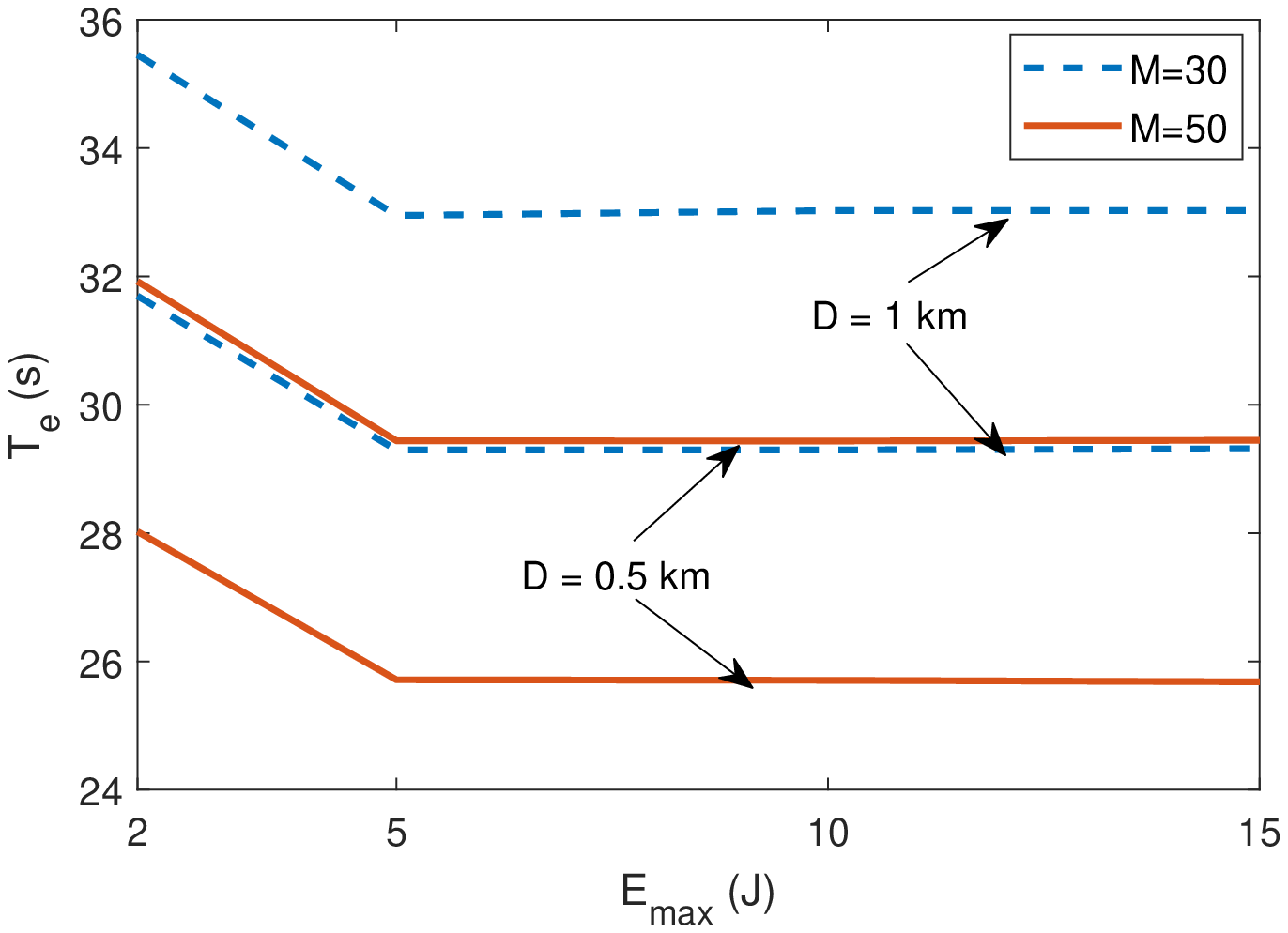}
\caption{Impact of UE's energy consumption limit $E_{\max}$ on the effective training time. Here, $K=4$.}
\label{fig:6}
\end{minipage}
\begin{minipage}[t]{0.4\textwidth}
\includegraphics[width=1\textwidth]{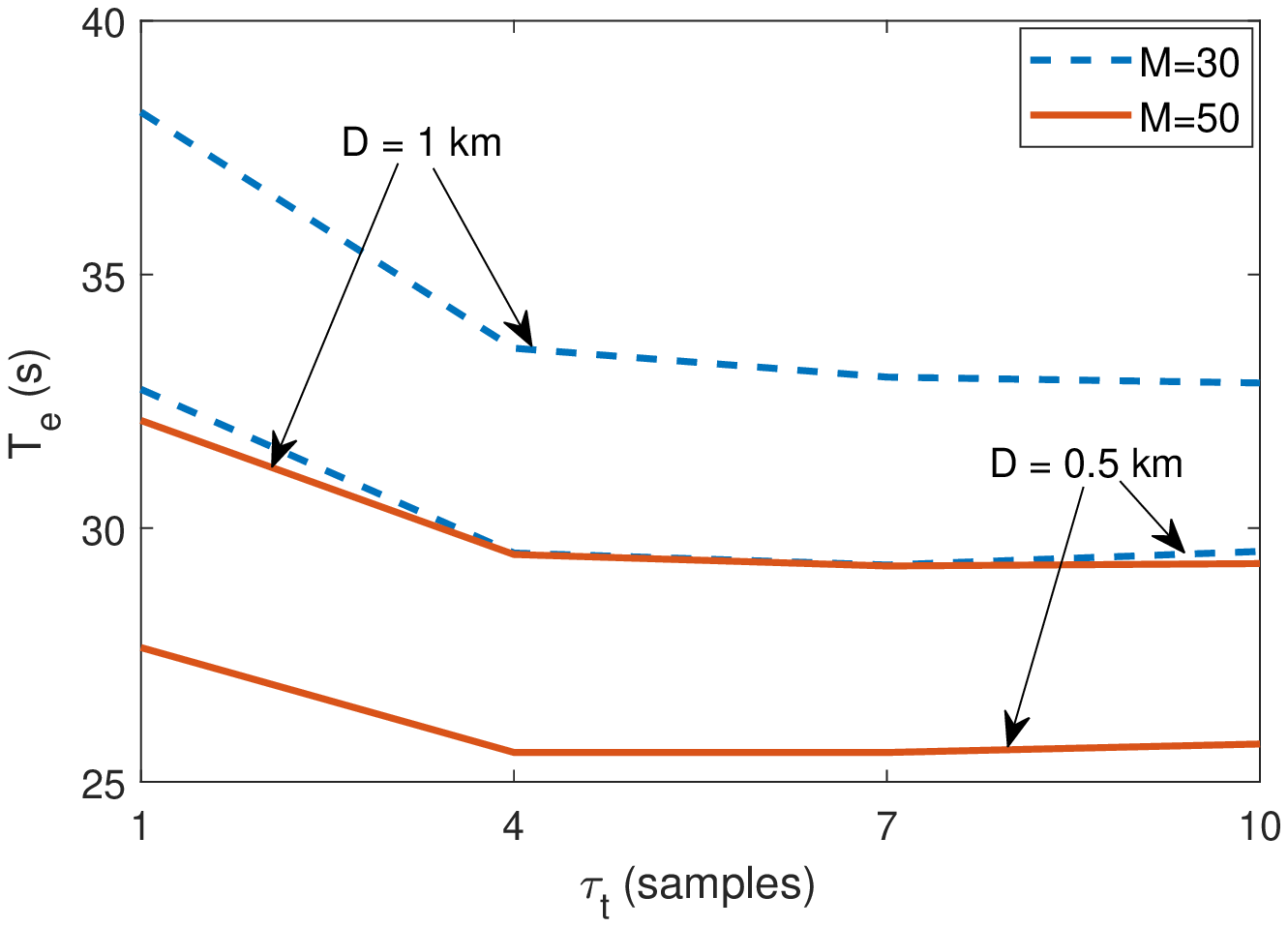}
\caption{Impact of the length of UL pilots on the effective training time. Here, $K=4$.}
\label{fig:7}
\end{minipage}
\end{figure}

In Fig.~\ref{fig:6}, the impact of UE's energy consumption limit $E_{\max}$ on the effective training time is revealed. Here, decreasing $E_{\max}$ leads to an increase in the effective time.
% but the time increase is not dramatic when the number of APs is small.
Specifically,
%when $M=50$ there is an increase of up to $7.1\%$ with $E_{\max}=2$ J, $D=0.5$ km in comparison to that with $E_{\max}=15$ J, $D=0.5$ km.
the increase is by up to $9.4\%$ with $E_{\max}=2$ J, $D=1$ km in comparison to that with $E_{\max}=15$ J, $D=1$ km.
%The increase is only up to $5\%$ with $E_{\max}=1$ J, $D=1$ km in comparison to that with $E_{\max}=15$ J, $D=1$ km.
This is reasonable because at a low value of $E_{\max}$, the effective time may not approach the optimal value due to a smaller feasible region of the optimization problem \eqref{mainP}.
We note that in the case of deep fading, the achievable rates of UEs could be small. This may lead to an infeasible problem because the constraint \eqref{cons:energy} on the energy consumed at each UE may be violated. However, it should be also emphasized that a cell-free massive MIMO network has many antennas distributed over a potentially large coverage area. As very high small-scale and macro diversity gains can be achieved, the probability of simultaneously experiencing deep fading for all links would be very small. The problem is therefore likely feasible and our proposed solution would apply.

Fig.~\ref{fig:7} focuses on the impact of the length of UL pilots on the effective training time. It is clear that too small and too large values of $\tau_t$ both increase the effective time.
% but the increase is not significant when the number of APs is large.
Specially, the effective time increases up to
%$6\%$ and $1\%$ with $\tau_t=1$ and $\tau_t=13$ in comparison to that with $\tau_t=7$, respectively. When $M=30$, the increases are up to
$10\%$ and $1\%$ with $\tau_t=1$ and $\tau_t=13$ in comparison to that with $\tau_t=7$, respectively.
This is reasonable because at a large value of $\tau_t$, the factor of $\frac{T_c-\tau_t}{T_c}$ makes the data rate decrease and the transmission time grows. In contrast, at a small value of $\tau_t$, the network suffers more from the pilot contamination, the data rates drop and the training update transmission time increases.
%\footnote{These paragraphs read boring because they have the same structure/rhythm...}
%In the latter case, the impact of the factor of $\frac{T_c-\tau_t}{T_c}$ is nominated by that of the pilot contamination.
%A higher number of APs leads to the larger feasible region of data rates, leading to a smaller change of the optimal value of the effective time.

\subsubsection{Cell-free massive MIMO vs. Cell-free TDMA massive MIMO}
\begin{figure}[t!]
\centering
\begin{minipage}[t]{0.4\textwidth}
\includegraphics[width=1\textwidth]{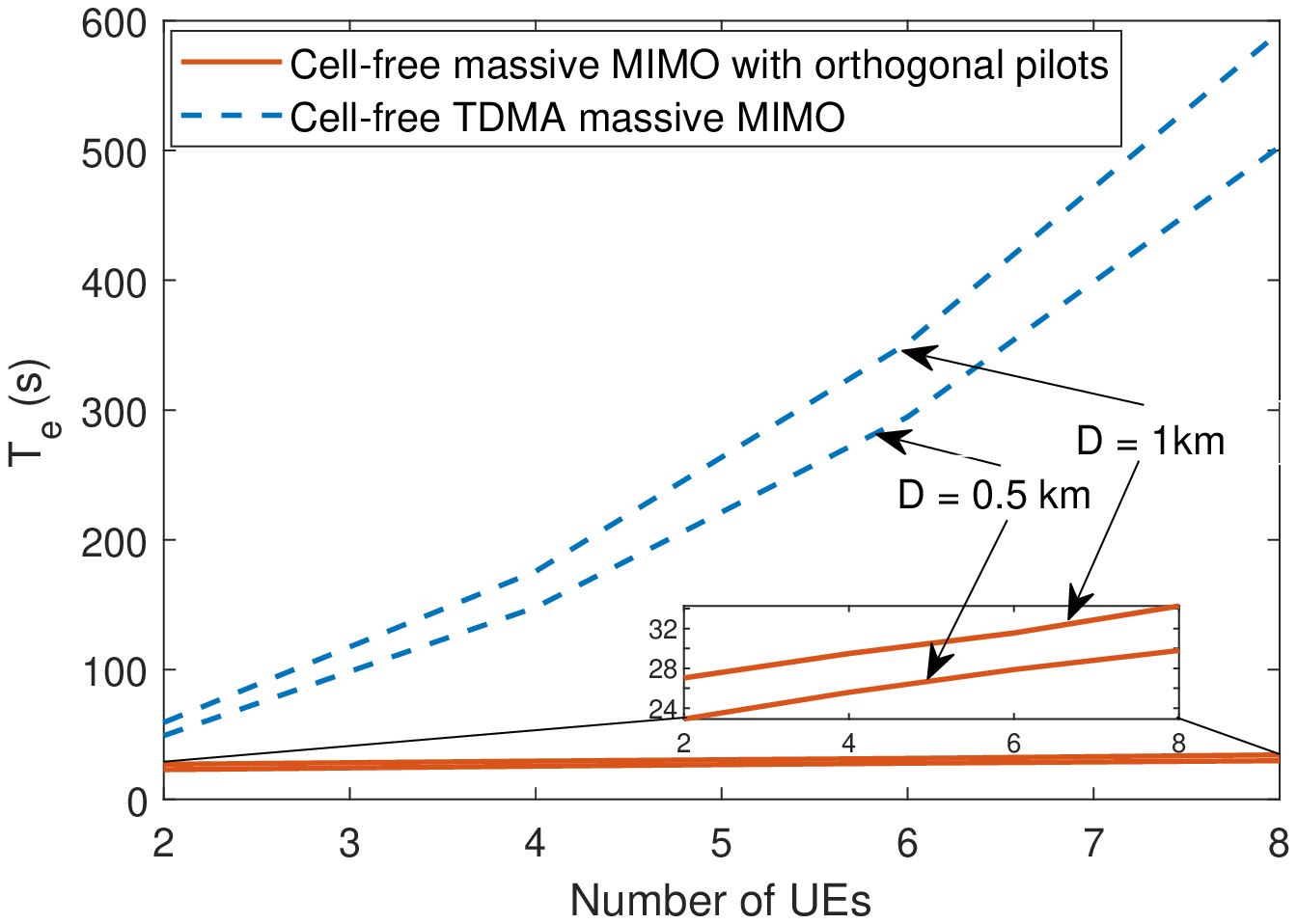}
\caption{Comparison between CFmMIMO with orthogonal pilots and cell-free TDMA massive MIMO. %with the comparison between the the cell-free TDMA massive MIMO network and the proposed CFmMIMO network.
Here, $M=50$.}
\label{fig:8}
\end{minipage}
\begin{minipage}[t]{0.4\textwidth}
\includegraphics[width=1\textwidth]{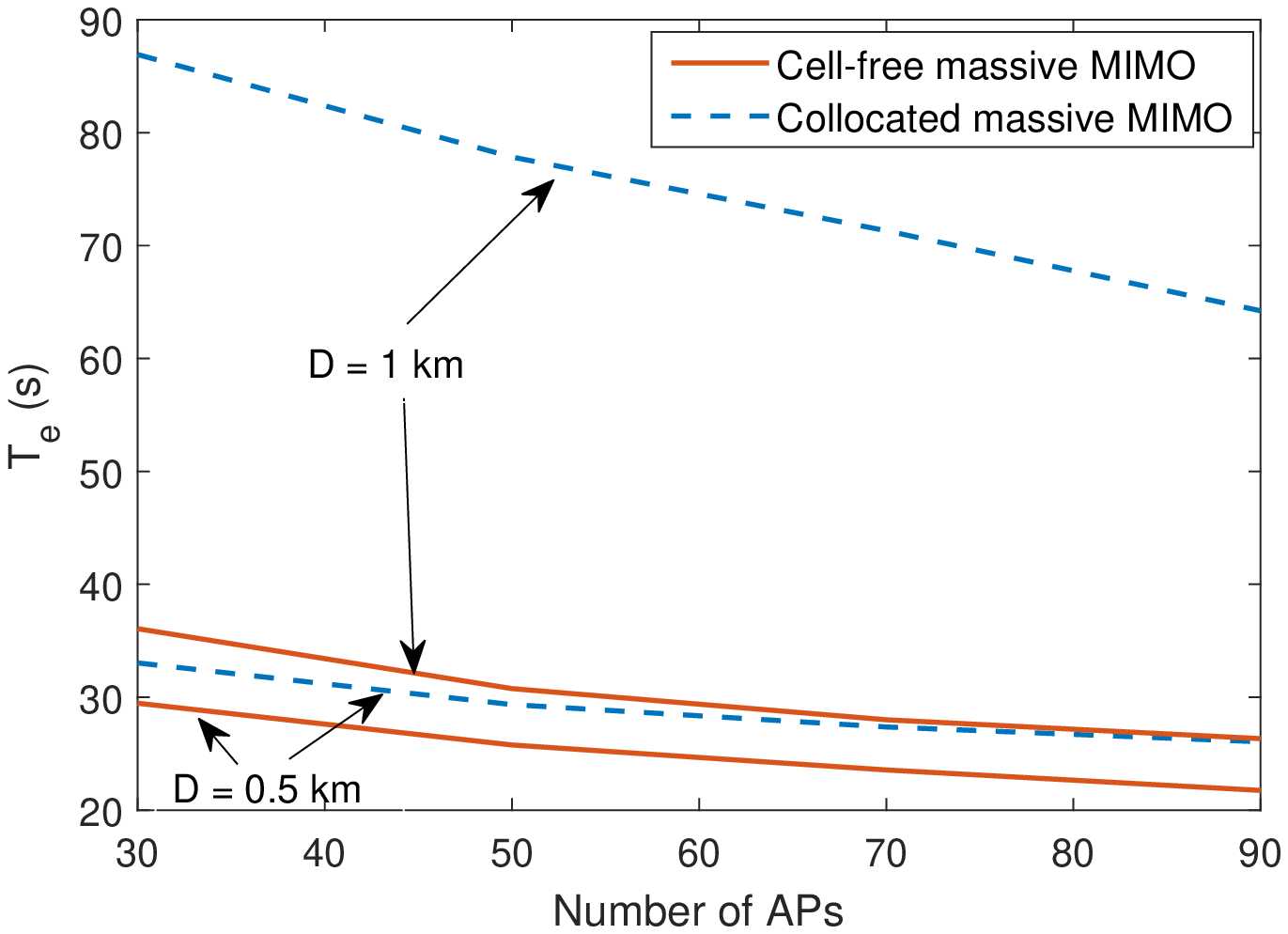}
\caption{Comparison between CFmMIMO and collocated massive MIMO. %with the comparison between the the CFmMIMO network and the collocated massive MIMO network.
Here, $K=4$.}
\label{fig:9}
\end{minipage}
\end{figure}

Fig.~\ref{fig:8} compares the cell-free massive MIMO system with the cell-free TDMA massive MIMO system. Since the pilot sequences of UEs in the latter are pairwisely orthogonal in the time domain, we choose orthogonal pilot sequences for the former, i.e., $\VARPHI_{\ell}^H\VARPHI_{k}=0, \forall \ell\in\K\setminus k$, for a fair comparison. The training durations are then the same for both systems.
We also choose $\widetilde{\rho}_t=K\rho_t$ for the amount of energy consumed at the ``UL channel estimation'' time blocks of the two networks to be the same, and $\tau_t=K$ so that the powers of channel estimate, i.e., $\sigma_{mk}^2,\forall m,k$, are the same in the two networks.
From Fig.~\ref{fig:8}, a significant time reduction (e.g., of up to $94\%$ with $K=8$) is achieved by the CFmMIMO compared with the cell-free TDMA massive MIMO. This result is expected because in the former, the factor of $(1/K)$ is imposed on the data rates and the training updates are transmitted sequentially. For a large number of UEs, the data rate is significantly small, and as a result, the training update transmission requires a substantially long time.

\subsubsection{Cell-free massive MIMO vs. Collocated massive MIMO}
Finally, we compare the effective training time in CFmMIMO  with that in collocated massive MIMO. Fig.~\ref{fig:9} shows that the former significantly outperform the latter, e.g., the time reduction is by up to $57\%$ with $M=30$ and $D=1$ km. This observation is as expected because CFmMIMO distributes antennas over their coverage area; and as such, their performance suffers less from the UEs with unfavorable links than that of collocated massive MIMO. Higher data rates and a lower training time then follow.
%From Figs.~\ref{fig:8} and~\ref{fig:9}, we conclude that in term of training time, CFmMIMO networks are best to support the considered FL framework.
%\end{remark}

\section{Conclusion}
\label{sec:con}
In this paper, we have proposed using CFmMIMO networks to support FL in a wireless environment. We designed a general scheme in which any algorithm and beamforming/filtering approach can be further developed to optimize the performance of any FL framework.
Specially here, each iteration of the FL optimization algorithms or the FL process
happens in one large-scale coherence time.
Targeting training time minimization for the FL framework \cite{ma15OMS} as example, we jointly design local accuracy, transmit power, data rate, and UE's processing frequency under the practical requirements on the UE's energy consumption limit and maximum transmit powers at the APs and UEs.
A mixed timescale stochastic nonconvex optimization problem has been formulated with the objective of minimizing the training time of one FL process. Based on the general online successive convex approximation framework, we have developed a new algorithm to successfully solve the formulated problem. We have proved that the proposed algorithm converges to the neighborhood of stationary points of the optimization problem.
%Each iteration of the algorithms happens only one large-scale coherence time.
For given parameter settings, numerical results show that our joint optimization design significantly reduces the training time of FL over the baselines under comparison.
%For a fast FL process, the values of local accuracy, UE's processing frequency, UE's energy usage, and the length of UL training pilots should be too small.
They have also confirmed that CFmMIMO offers the lowest training time when compared with cell-free TDMA massive MIMO and collocated massive MIMO. %Therefore, they are potentially promising solutions to support FL in future wireless communications networks.
\ifCLASSOPTIONcaptionsoff
  \newpage
\fi

\bibliographystyle{IEEEtran}
\bibliography{IEEEabrv,newidea2019}
\vspace{-0mm}
\begin{IEEEbiography}
%[{\includegraphics*[width=1in, height=1.25in, clip, keepaspectratio]{tung}}]
{Tung Thanh Vu} (S'17) received the B.Sc. degree (Hons.) in telecommunications and networking from Ho Chi Minh City University of Science in 2012 and the M.Sc. degree (Hons.) in telecommunications engineering from Ho Chi Minh City University of Technology in 2016. He is currently a PhD candidate in the School of Electrical Engineering and Computing, The University of Newcastle, Australia. His current research interests include optimization designs and machine learning for cell-free massive MIMO and cloud radio access networks.
\end{IEEEbiography}
\vspace{-0mm}
\begin{IEEEbiography}
%[{\includegraphics*[width=1in, height=1.25in, clip, keepaspectratio]{ngo}}]
{Duy Trong Ngo} (S'08-M'15) received the B.Eng. degree (Hons.) in telecommunication engineering from The University of New South Wales, Australia, in 2007, the M.Sc. degree in electrical engineering (communication) from the University of Alberta, Canada, in 2009, and the Ph.D. degree in electrical engineering from McGill University, Canada, in 2013.
He is a Senior Lecturer with the School of Electrical Engineering and Computing, The University of Newcastle, Australia, where he is currently involved in the research effort of design and optimization for 5G and beyond wireless communications networks. His current research interests include cloud radio access networks, multi-access edge computing, and vehicle-to-everything (V2X) communications for intelligent transportation systems.
\end{IEEEbiography}
\vspace{-0mm}
\begin{IEEEbiography}
%[{\includegraphics[width=1in,height=1.25in,clip, keepaspectratio]{nguyen}}]
{Nguyen H. Tran} (S'10-M'11-SM'18) received BS and Ph.D degrees, from HCMC University of Technology and Kyung Hee University, in electrical and computer engineering, in 2005 and 2011, respectively. He was an Assistant Professor with Department of Computer Science and Engineering, Kyung Hee University, from 2012 to 2017. Since 2018, he has been with the School of Computer Science, The University of Sydney, where he is currently a Senior Lecturer. His research interests include distributed computing, machine learning, and networking. He received the best KHU thesis award in engineering in 2011 and several best paper awards, including IEEE ICC 2016, APNOMS 2016, IEEE ICCS 2016, and ACM MSWiM 2019. He receives the Korea NRF Funding for Basic Science and Research 2016-2023 and ARC Discovery Project 2020-2023. He has been the Editor of IEEE Transactions on Green Communications and Networking since 2016.
\end{IEEEbiography}
\vspace{-0mm}
\begin{IEEEbiography}
%[{\includegraphics[width=1in,height=1.25in,clip, keepaspectratio]{hien}}]
{Hien Quoc Ngo} is currently a Lecturer at Queen's University Belfast, UK. His main research interests include massive (large-scale) MIMO systems, cellfree massive MIMO, physical layer security, and cooperative communications. He has co-authored
many research papers in wireless communications and co-authored the Cambridge University Press textbook Fundamentals of Massive MIMO (2016). He received the IEEE ComSoc Stephen O. Rice Prize in Communications Theory in 2015, the IEEE ComSoc Leonard G. Abraham Prize in 2017, and the Best Ph.D. Award from EURASIP in 2018.
\end{IEEEbiography}
\vspace{-0mm}
\begin{IEEEbiography}
%[{\includegraphics[width=1in,height=1.25in]{minh}}]
{Minh Ngoc Dao} received the Ph.D. degree in applied mathematics from the University of Toulouse, France, in 2014. He was a Postdoctoral Fellow with The University of British Columbia, Canada, from 2014 to 2016, and a Research Associate with The University of Newcastle, Australia, from 2016 to 2019. He is currently a Research Associate with The University of New South Wales, Australia. His research interests include nonlinear optimization, nonsmooth analysis, control theory, signal processing, and machine learning. In 2017, he received the Annual Best Paper Award from the Journal of Global Optimization.
\end{IEEEbiography}
\vspace{-0mm}
\begin{IEEEbiography}
%[{\includegraphics[width=1in,height=1.25in]{rick}}]
{Richard H. Middleton} (SM'86–-F'99) completed his Ph.D. (1987) from the University of Newcastle, Australia. He was a Research Professor at the Hamilton Institute, The National University of Ireland, Maynooth from May 2007 till 2011 and is currently Professor at the University of Newcastle and Head of the School of Electrical Engineering and Computing. He has served as Program Chair (CDC 2006), co-general chair (CDC 2017) CSS Vice President Membership Activities, and Vice President Conference Activities. In 2011, he was President of the IEEE Control Systems Society. He is a Fellow of IEEE and of IFAC, and his research interests include a broad range of Control Systems Theory and Applications, including Communications Systems, control of distributed systems and Systems Biology.
\end{IEEEbiography}
%\balance
\end{document}